\documentclass[fleqn,usenatbib]{mnras}
\usepackage{newtxtext,newtxmath}

\usepackage[T1]{fontenc}
\usepackage{ae,aecompl}

\usepackage{graphicx}	
\usepackage{amsmath}	
\usepackage{amssymb}	
\usepackage{times}
\usepackage{siunitx}

\newcommand{\kms}{\si{km.s^{-1}}}
\newcommand{\doh}{$\Delta$(O/H)}
\newcommand{\sigmaoh}{$\sigma$(O/H)}
\newcommand{\hii}{\ion{H}{II}}

\usepackage{etoolbox}
\makeatletter
\patchcmd\@combinedblfloats{\box\@outputbox}{\unvbox\@outputbox}{}{%
    \errmessage{\noexpand\@combinedblfloats could not be patched}%
}%
\makeatother

\title[Measuring the mixing scale of the ISM]{Measuring the mixing scale of the ISM within nearby spiral galaxies}

\author[K. Kreckel et al.]
{Kathryn Kreckel$^{1}$\thanks{E-mail: kathryn.kreckel@uni-heidelberg.de},
I-Ting Ho$^{2}$, 
Guillermo A. Blanc$^{3,4}$,
Simon C.~O.~Glover$^{5}$, 
\newauthor
Brent Groves$^{7}$,
Erik Rosolowsky$^{8}$, 
Frank Bigiel$^{9}$,
M\'{e}d\'{e}ric Boqu\'{i}en$^{10}$, 
\newauthor
M\'elanie~Chevance$^{1}$,
Daniel A. Dale$^{11}$,
Sinan Deger$^{12}$,
Eric Emsellem$^{13,14}$,
\newauthor
Kathryn Grasha$^{15}$,
Jenny J. Kim$^{1}$, 
Ralf S. Klessen$^{5,6}$,
J.~M.~Diederik Kruijssen$^{1}$,
\newauthor
Janice C. Lee$^{12}$,
Adam K.~Leroy$^{16}$,
Daizhong Liu$^{2}$,
Rebecca McElroy$^{17}$,
\newauthor
Sharon E.~Meidt$^{18}$,
Ismael Pessa$^{2}$,
Patricia Sanchez-Blazquez$^{19}$,
Karin Sandstrom$^{20}$,
\newauthor
Francesco~Santoro$^{2}$,
Fabian Scheuermann$^{1}$,
Eva Schinnerer$^{2}$,
Andreas Schruba$^{21}$,
\newauthor
Dyas Utomo$^{16}$,
Elizabeth J. Watkins$^{1}$,
Thomas G. Williams$^{2}$
}

\date{Accepted XXX. Received YYY; in original form ZZZ}

\pubyear{2020}

\begin{document}
\label{firstpage}
\pagerange{\pageref{firstpage}--\pageref{lastpage}}
\maketitle
I 
\begin{abstract}
The spatial distribution of metals reflects, and can be used to constrain, the processes of chemical enrichment and mixing. Using PHANGS-MUSE optical integral field spectroscopy, we measure the gas phase oxygen abundances (metallicities) across 7138 \hii\ regions in a sample of eight nearby disc galaxies.  In Paper I \citep{Kreckel2019} we measure and report linear radial gradients in the metallicities of each galaxy, and qualitatively searched for azimuthal abundance variations.  Here, we examine the two-dimensional variation in abundances once the radial gradient is subtracted, \doh, in order to quantify the homogeneity of the metal distribution and to measure the mixing scale over which HII region metallicities are correlated.
We observe low (0.03--0.05 dex) scatter in \doh\ globally in all galaxies, with significantly lower (0.02--0.03 dex) scatter on small ($<$600~pc) spatial scales.  This is consistent with the measurement uncertainties, and implies the two-dimensional metallicity distribution is highly correlated on scales of $\lesssim 600$"~pc. We compute the two point correlation function for metals in the disc in order to quantify the scale lengths associated with the observed homogeneity. This mixing scale is observed to correlate better with the local gas velocity dispersion (of both cold and ionized gas) than with the star formation rate.  Selecting only \hii\ regions with enhanced abundances relative to a linear radial gradient, we do not observe increased homogeneity on small scales. This suggests that the observed homogeneity is driven by the mixing introducing material from large scales rather than by pollution from recent and on-going star formation.
\end{abstract}




\section{Introduction}
Dynamical conditions within the interstellar medium (ISM) play an important role in regulating galaxy evolution.  The cloud-scale turbulent state \citep{Larson1981} combined with the surrounding ISM pressure \citep{Hughes2013, Schruba2019, Sun2020} and large-scale dynamical processes \citep{Colombo2014,Jeffreson2018,Meidt2018,Meidt2020} set the conditions for cloud stability and cloud collapse, regulating future star formation.   Cloud collapse is additionally regulated by the physical and chemical conditions in the ISM, as the presence of heavy elements  facilitates gas cooling \citep{Dopita2003,Smith2017, Klessen2016}. 
As a result, understanding the homogeneity (or inhomogeneity) in the gas-phase abundances (metallicity) is important for informing our understanding of the star formation process.  The distribution of elements in the gas phase has further implications for star formation history studies based on observed stellar metallicities, such as Galactic archaeology and (globular) cluster formation, which presuppose a certain level of homogeneity \citep[see review in][]{Krumholz2019}. 

Many processes are expected to drive mixing of metals within the ISM, increasing the homogeneity. This includes bar-driven radial mixing \citep{DiMatteo2013}, spiral arm driven large-scale systematic streaming motions \citep{Grand2016, Sanchez-Menguiano2016}, kiloparsec-scale mixing-induced dilution due to the spiral density waves passage \citep{Ho2017}, thermal and gravitational instabilities \citep{Yang2012, Petit2015}, and interstellar turbulence \citep{deAvillez2002,Klessen2003, Krumholz2018}. Many of these processes derive from non-axisymmetric features like bars and spiral arms, and are expected to preserve and/or introduce azimuthal variations in the metallicity distribution.  To observationally distinguish these models, we need to quantify the level of homogeneity,  the spatial scales over which mixing is effective, and how it relates to large-scale dynamics and environments. 

\hii\ regions, photoionized by the most massive OB stars during their short ($<$10 Myr) lifetimes, exhibit line emission that probes the instantaneous local ISM abundances. 
With a sufficiently large sample of \hii\ regions, it is possible to use these as tracers of the underlying two dimensional metallicity field, mapping the effectiveness of mixing in the ISM across a galaxy disc.  Until the last decade, \hii\ region metallicity studies were limited to samples of tens of regions per galaxy \citep{Pilyugin2014}, sufficient only for measuring the radial metallicity gradient.  Long-slit multi-object spectroscopy has achieved samples of $\sim$100 \hii\ regions per galaxy \citep{Berg2015, Croxall2015, Croxall2016, Berg2020}, however in recent years, optical integral field unit (IFU) spectroscopic techniques have increased these numbers to cover 1000s of \hii\ regions within individual galaxies \citep{Rosales-Ortega2011, Sanchez-Menguiano2016, Ho2017}, achieving the necessary statistics to investigate azimuthal metallicity variations.  

Mapping both the radial and azimuthal variations in metallicity across a sample of galaxies is a key science goal of the PHANGS\footnote{Physics at High Angular resolution in Nearby GalaxieS; \url{http://www.phangs.org}} collaboration. With our PHANGS-MUSE large observing program (PI:Schinnerer; 1100.B-0651), we use the Very Large Telescope/Multi Unit Spectroscopic Explorer (VLT/MUSE) to mosaic the central 4--8~kpc disc of 19 galaxies with optical IFU observations. Preliminary results from the first eight completed galaxies in the PHANGS-MUSE survey (\citealt{Kreckel2019}, hereafter Paper I) reveal that half exhibit significant azimuthal variations, and suggest that spiral arms play a role in organizing and mixing the ISM. These statistics have been recently confirmed with a larger sample of 45 disc galaxies \citep{SanchezMenguiano2020}.

Building on Paper I, we use our sample of 7,138 \hii\ regions across the first eight PHANGS-MUSE galaxies (Section \ref{sec:data}) to quantify the chemical homogeneity of the ISM (Section \ref{sec:homogeneity}) and measure the characteristic physical scale below which the gas-phase ISM is chemically homogeneous (Section \ref{sec:corr}). In Section \ref{sec:discussion}, we compare these results with theoretical models, and relate them to the spiral structure of the host galaxy, before concluding in Section \ref{sec:conclusion}.

\section{Data}
\label{sec:data}
\begin{table*}
\footnotesize
\caption{Key parameters of our galaxy sample, adopted from the z0mgs (z=0 multiwavelength galaxy survey; \citealt{Leroy2019}) database (\url{https://github.com/akleroy/z0mgs}), where the original references can be found.} \label{tab:sample}
\begin{tabular}{lccrrrrcc}
\hline
{Name} &
{D} &
{Type} &
{Log$_{10}$( $\frac{{\rm M}_*}{{\rm M}_{\odot}}$ )} & 
{Log$_{10}$( $\frac{{\rm SFR}}{{\rm M}_{\odot}{\rm yr}^{-1}}$ )} & 
{PA} &
{i} &
{Physical} &
{Number of}  
\\
 &
{} &
{} &
{} & 
{} & 
{} &
{} &
{resolution} &
{HII regions}  
\\
{} &
{(Mpc)} &
{} &
{} & 
{} & 
{(degrees)} &
{(degrees)} &
{(pc)} &
{} 
\\
\hline
NGC\,0628 & $9.77\pm0.82$ & Sc & 10.2 & 0.26 & 20.9 & 8.7    & 47 & 1277  \\
NGC\,1087 & $14.4\pm4.8$ & SBc & 9.8 & 0.05 & 177.3 & 41.3 &   70 & 679 \\
NGC\,1672 & $11.9\pm4.0$ & SBb (S) & 10.2 & 0.48 & 135.7 & 37.5 &   46 & 880 \\
NGC\,2835 & $10.1\pm3.4$ & SBc & 9.6 & -0.08 & 1.6 & 47.8 &  44 & 699 \\
NGC\,3627 & $10.6\pm0.9$ & SBb (S3) & 10.6 & 0.55 & 174.1 & 55.0 &   51 & 692 \\ 
NGC\,4254 & $16.8\pm5.6$ & Sc & 10.5 & 0.74 & 67.7  & 37.8  & 57 & 1824 \\ 
NGC\,4535 & $15.8\pm2.3$ & SBc & 10.4 & 0.35 & 179.8  &    40.7 &  38 &  864 \\ 
IC\,5332 &  $9.95\pm3.3$  & SBc & 9.6 & -0.31 & 74.7  &    24.0   & 39 &  223  \\
\hline
\end{tabular}
\end{table*}

\begin{figure*}
\centering
\includegraphics[height=8.5in]{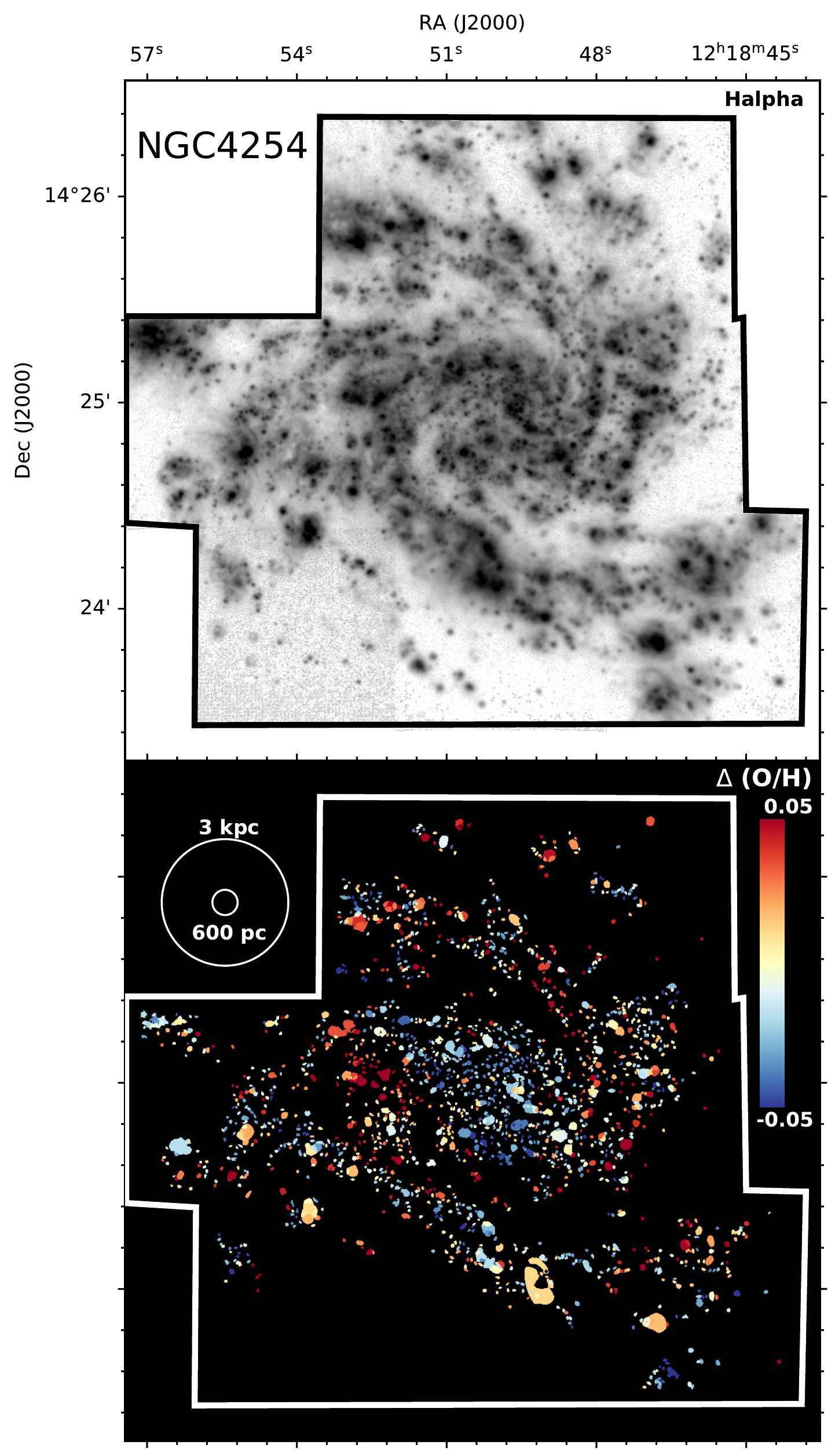}
\caption{Mapping the \hii\ regions in NGC~4254.  Top: H$\alpha$ emission is detected across the disk of the galaxy. In Paper I we describe how the H$\alpha$ morphology is used to identify \hii\ region boundaries, resulting in a sample of \hii\ regions that have been isolated from the surrounding diffuse ionized gas. Bottom: Color coding each \hii\ region by its metallicity offset from a radial gradient, \doh, it is clear that we sample the gas phase oxygen abundances throughout the disk.  In this paper, we quantify the structures revealed in these \doh\ maps.   
\label{fig:ngc4254}}
\end{figure*}

For this work, we analyse results from the first eight galaxies observed with the MUSE spectrograph \citep{Bacon2010} at the VLT as part of the PHANGS-MUSE survey (Table \ref{tab:sample}; Paper I).  In general, PHANGS galaxies have been selected to be nearby (D$<$17 Mpc), low inclination (i $<$ 60$^\circ$) main-sequence disc galaxies.  These first eight galaxies span a range in stellar mass of only one order of magnitude, such that they all have approximately solar or slightly sub-solar metallicities.  

All galaxies are at sufficiently small distances such that the physical resolution is better than 70~pc given the $\sim$1\arcsec PSF sizes. This sets the minimum \hii\ region separation length to which we are sensitive. Median \hii\ region separations within our catalogues range from 100--150~pc, and median \hii\ region diameters are $\sim$80~pc (marginally resolved). 
All physical size scales throughout this paper are calculated based on deprojected separations, assuming the galaxy distance (D), position angle (PA) and inclination (i) listed in Table \ref{tab:sample}. 

Images of all targets and associated \hii\ region catalogues, along with details on the observing setup, data reduction, techniques for selecting \hii\ regions and prescriptions applied to infer the gas phase oxygen abundances are described in Paper I, and only briefly summarized here.

\subsection{HII region metallicities}
Using the Balmer decrement, assuming case B recombination and $T_{\rm e}=10^4$~K, we apply the \cite{Fitzpatrick1999} extinction law assuming a value of $R_V=3.1$ in order to determine de-reddened line flux measurements for all strong emission lines (H$\beta$, [OIII] $\lambda$4959,5007, [NII]$\lambda$6548,6583, H$\alpha$, [SII]$\lambda$6716,6731). All metallicity calculations  in this paper adopt the empirical \cite{Pilyugin2016} S-calibration (Scal).   
It relies on the following three standard diagnostic line ratios: \\ 
$N_2$  = $I_{\rm [N\,II] \lambda 6548+ \lambda 6584} /I_{{\rm H}\beta }$,  \\
$S_2$  = $I_{\rm [S\,II] \lambda 6717+ \lambda 6731} /I_{{\rm H}\beta }$,  \\
$R_3$  = $I_{{\rm [O\,III]} \lambda 4959+ \lambda 5007} /I_{{\rm H}\beta }$.  \\
Here we have measured only the stronger line in both the [OIII] and [NII] doublets and assume a fixed ratio of 3:1 \citep{Storey2000}.  The Scal prescription is defined separately over the upper and lower branches in log$N_{2}$, with almost all \hii\ regions in our sample falling on the upper branch
(log$N_{2} \ge -0.6$), where
\begin{eqnarray}
\footnotesize
       \begin{array}{lll}
     {\rm 12+log(O/H)}  & = &   8.424 + 0.030 \, \log (R_{3}/S_{2}) + 0.751 \, \log N_{2}   \\  
                          & + &  (-0.349 + 0.182 \, \log (R_{3}/S_{2}) + 0.508 \log N_{2})   \\ 
                          & \times & \log S_{2}.   \\ 
     \end{array}
\label{equation:ohsu}
\end{eqnarray}
For the lower branch
(log$N_{2} \le -0.6$), the oxygen abundance is instead calculated as
\begin{eqnarray}
\footnotesize
       \begin{array}{lll}
     {\rm 12+log(O/H)}  & = &   8.072 + 0.789 \, \log (R_{3}/S_{2}) + 0.726 \, \log N_{2}   \\  
                          & + &  (1.069 - 0.170 \, \log (R_{3}/S_{2}) + 0.022 \log N_{2})    \\ 
                          & \times & \log S_{2}.   \\ 
     \end{array}
\label{equation:ohsl}
\end{eqnarray}
 In Appendix \ref{appendix}, we repeat our calculations using other common metallicity calibrations. These yield qualitatively similar results to our preferred Scal method.  \cite{Ho+2019} used auroral line detections to directly measure abundance variations for one galaxy in our sample, NGC~1672,  across \hii\ regions in one spiral arm.  Those authors find very good qualitative agreement with the trends identified using the Scal strong line method, increasing our confidence in the use of this calibration, particularly in relation to relative metallicity trends. 

Typical uncertainties in the measured metallicities, based on propagation of systematic errors in the data reduction and random errors associated with the line flux measurement, are 0.02 dex. Because our analysis compares variations in metallicity, constant offsets due to calibration uncertainties are effectively removed from the analysis.  

In our results, we refer to the metallicity variation once the radial gradient is subtracted as \doh (see Paper I).   Subtracting a median radial gradient instead of a linear fit does not change the results presented here.  We refer to the 1$\sigma$ scatter observed in \doh\ as $\sigma$(O/H).  Both \doh\ and \sigmaoh\ are reported in logarithmic abundance units (e.g. log(O/H)).  Figure \ref{fig:ngc4254} shows the H$\alpha$ map and selected \hii\ regions within NGC~4254, color coded by \doh\, as an example.

\subsection{Ancillary data}
\label{sec:ancillary}
In attempting to understand the physical mechanisms affecting our local metallicity variations, we compare our results to a variety of physical conditions inferred from ancillary data. As many parameters vary radially, and disc rotation impacts the azimuthal mixing of the ISM, we choose to do this in radial bins.  For our analysis, we exclude the central 1~kpc and break down the remaining area of our galaxies into 2~kpc wide annuli. This size scale is chosen to ensure a sufficiently large sample  of \hii\ regions ($\sim$100) in most annuli, and is larger than the poorest angular resolution available in our ancillary data (\ion{H}{i} maps at 25\arcsec). Results presented here are not sensitive to the width of the annuli.

A natural potential source of turbulence and mixing is the mechanical and chemical feedback from the star formation processes itself within any given annulus.  To calculate the star formation rate (SFR) and SFR surface density ($\Sigma_{\rm SFR}$), we sum the extinction-corrected \hii\ region H$\alpha$ luminosities, L(H$\alpha$), across the annulus.  We then convert to SFR following \cite{Murphy2011} as
\begin{equation}
\mathrm{SFR}(\mathrm{M}_\odot~\mathrm{yr^{-1}}) = 5.37 \times 10^{-42} L(\mathrm{H}\alpha) (\si{erg.s^{-1}})
\end{equation} 
which assumes a constant star formation rate over 100~Myr and a fully-sampled Kroupa IMF \citep{Kroupa2001}. This should be a reasonable approximation for our 2-kpc wide rings, which each cover a large part of the galaxy. Our results do not change qualitatively if we include the diffuse H$\alpha$ emission within the annulus in this analysis.

The orbital time within any given annulus is determined from high resolution rotation curve fits to the molecular gas kinematics, as traced by PHANGS-ALMA observations of CO (2-1) emission \citep{Lang2020}.  IC~5332 is excluded from this analysis, as it contains no detected pixel above the necessary S/N cut. As the molecular gas is largely confined to a thin mid-plane disc, we expect the derived rotation curve to serve as a reasonable tracer of the orbital time experienced by the  \hii\ regions.

Accounting for the molecular and atomic gas relative to the stellar mass, we measure the gas fraction in each annulus. All three tracers are currently available for only six of our eight galaxies, as IC~5332 and NGC~1672 currently have no \ion{H}{i} observations available and are thus excluded. The total gas fraction f$_g$ is then calculated as
\begin{equation}
f_g=\frac{\Sigma (M_{\rm H_2})+\Sigma (M_{\rm HI})}{\Sigma (M_{\rm H_2})+\Sigma (M_{\rm HI})+\Sigma (M_*)}.
\label{eqn:fg}
\end{equation}

The stellar mass is mapped directly from the MUSE data cubes (Pessa et al. in prep).  Voronoi bins are constructed with a target continuum S/N of 35, and the stellar population is fit using the Penalized Pixel-Fitting method (pPXF; \citealt{Cappellari2004,Cappellari2017}) without regularization, and masking of all emission lines and strong sky lines.  As input templates for pPXF we take the eMILES stellar population (SSP) models \citep{Vazdekis2010}  using a Kroupa IMF, Basti isochrones, and base (matching the Galactic abundances) $\alpha$/Fe. Stellar masses are then calculated by taking into account the mass to light ratio of the stellar SSP templates.  With this approach, we find typical agreements within 40\% to estimates derived from  mid-infrared stellar emission (e.g. \citealt{Querejeta2015}), where a constant M/L ratio was assumed after removing the non-stellar emission. This difference has proved to be strongly correlated with the age of the underlying stellar population in a given spaxel, where the overestimation of the mass value from mid-IR emission with respect to MUSE is more severe in spaxels dominated by younger stellar populations. Quereta et al in prep will address these differences more in depth.

The molecular gas mass is measured from the PHANGS-ALMA survey CO(2-1) moment-zero maps at 1-2\arcsec\ resolution (Leroy et al. in prep), assuming a fixed CO(2-1)/CO(1-0) ratio of 0.7 (\citealt{Leroy2013}; den Brok et al. subm, Saito et al. in prep) and a Milky Way CO-to-H$_{2}$ conversion factor of 4.4~M$_\odot$~pc$^{-2}$/(K~km~s$^{-1}$) \citep{Bolatto2013}, which includes a factor of 1.36 to account for helium. Most of our targets are approximately Milky Way mass galaxies at solar metallicity, making this a reasonable assumption.  Choosing to adopt a metallicity-dependent conversion factor \citep{Bolatto2013} would result in a $\sim$20\% increase to the molecular gas estimates. Molecular gas maps for six of our eight targets have been reported by our earlier works \citep{Sun2018, Kreckel2018, Utomo2018}, and we refer the reader to those papers for additional visualizations and details. 

The atomic gas mass is measured from 21~cm \ion{H}{i} emission observed at the Karl G. Jansky Very Large Array (VLA) as part of archival (THINGS, \citealt{Walter2008}; VIVA, \citealt{Chung2009}) and new data (Legacy ID: AU157, PI: D. Utomo), and includes a correction for helium.  The angular resolution ranges from 15\arcsec\ to 25\arcsec, well matched to the $\sim$2~kpc scales considered in our annuli for the most distant galaxies.

The H$\alpha$ velocity dispersion ($\sigma_{\rm H\alpha}$, deconvolved from the instrumental response, but not corrected for thermal broadening) is measured within each \hii\ region. We note, however, that the MUSE spectral resolution ($\sim$49 \kms at H$\alpha$; \citealt{Bacon2017}) is much larger than our measured values (10--40 \kms), and without an extremely careful determination of the line spread function we caution against placing high confidence in these measurements. However, given the exceptionally high S/N in H$\alpha$ within the \hii\ regions, this provides the most robust determination, and we take $\sigma_{\rm H\alpha}$ to be the median value of all \hii\ regions within the annulus. As a test, we calculated the median H$\alpha$ velocity dispersion using all pixels that fall within the annulus, which results in $\sim$5 \kms\ higher H$\alpha$ velocity dispersions. This is likely due to either the increased turbulence or the higher scale height in the diffuse ionized gas \citep{DellaBruna2020}.   

The CO velocity dispersion ($\sigma_{\rm CO}$) is measured from PHANGS-ALMA data \citep[line widths for the PHANGS-ALMA data have been presented and investigated in][]{Sun2018,Sun2020}.  Here we use the velocity dispersion estimated from the rms scatter of emission about the intensity-weighted mean velocity (the ``moment 2'' line width). Use of an alternate parameterization of the molecular gas line width does not significantly change our results. As this emission predominantly arises from within cold molecular clouds, it shows systematically lower velocity dispersions than the ionized gas. 



\section{Quantifying the chemical homogeneity of the ISM}
\label{sec:homogeneity}

\begin{figure*}
\centering
\includegraphics[height=2in]{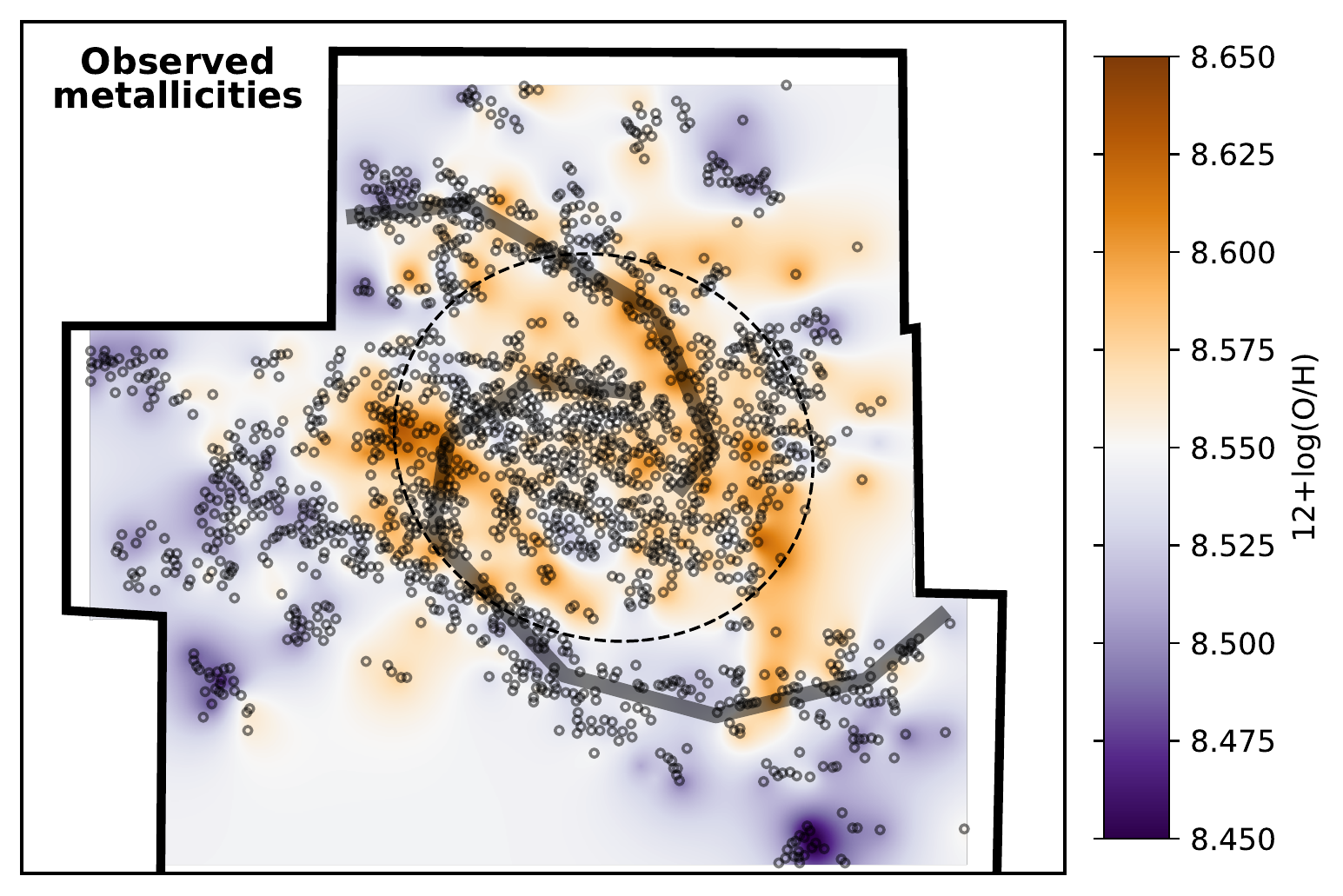}
\includegraphics[height=2in]{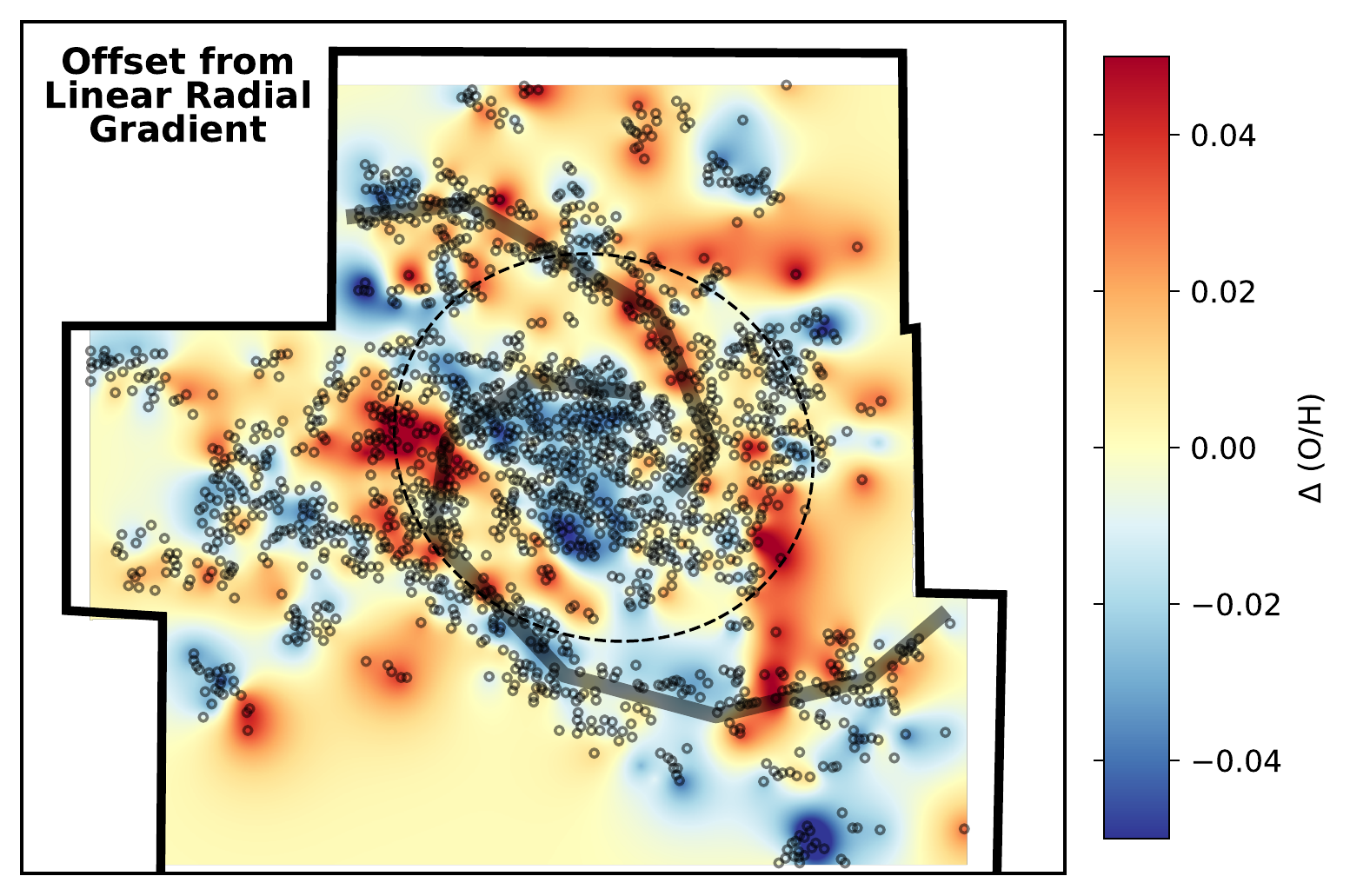}
\includegraphics[height=2in]{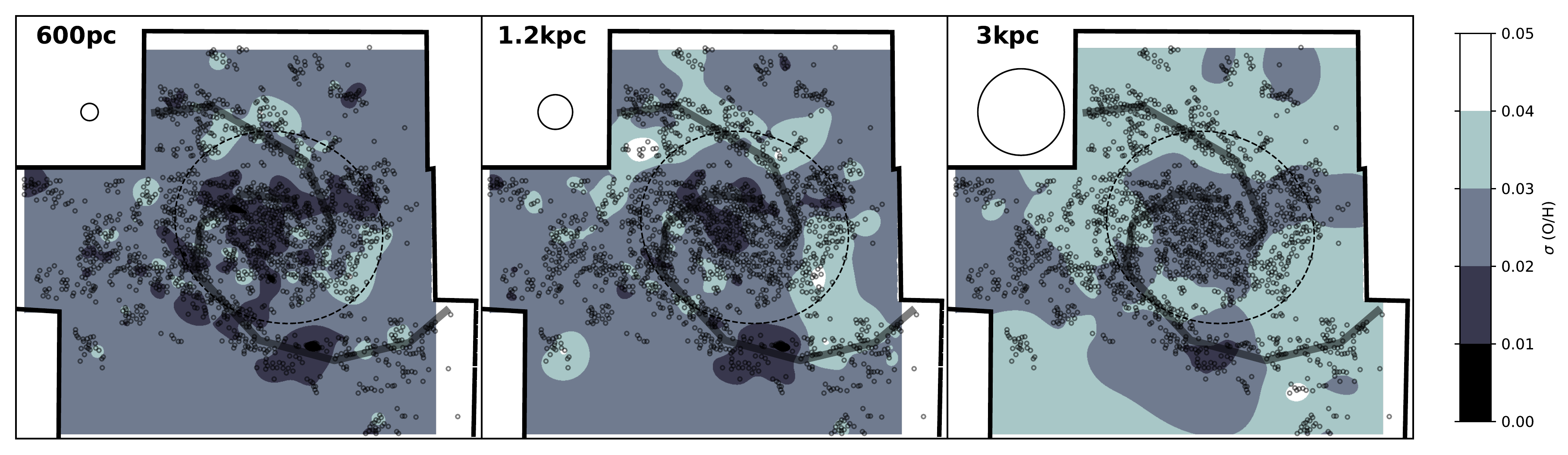}
\caption{Visualizing the large-scale homogeneity in metallicities within NGC~4254.  Circles represent individual \hii\ regions.  The dashed circle is at $0.3R_{25}$, and grey lines roughly outline the spiral structure. In colorscale we show an interpolation between the values using {\tt kriging}, a local estimation technique that derives the  best linear unbiased prediction of the intermediate values.  This is done purely for visualization purposes, and assumes an exponential model with 100~pc scale length, which is approximately twice our spatial resolution (see Table \ref{tab:sample}).   We map the observed metallicity (12+log(O/H), top left) and the offset from a linear radial gradient (\doh, top right; see Paper I).  For the bottom panels, we have calculated the standard deviation of metallicity values between each HII region and its neighbors over increasing distances (600~pc, 1.2~kpc and 3~kpc diameters).  We see systematically reduced scatter on the smallest 600~pc spatial scales.  This is also visible in the maps of \doh, where this $\sim$600~pc scale is well matched to the scale length over which uniform values are observed.  
\label{fig:visualize_new}}
\end{figure*}

\begin{figure}
\centering
\includegraphics[width=3in]{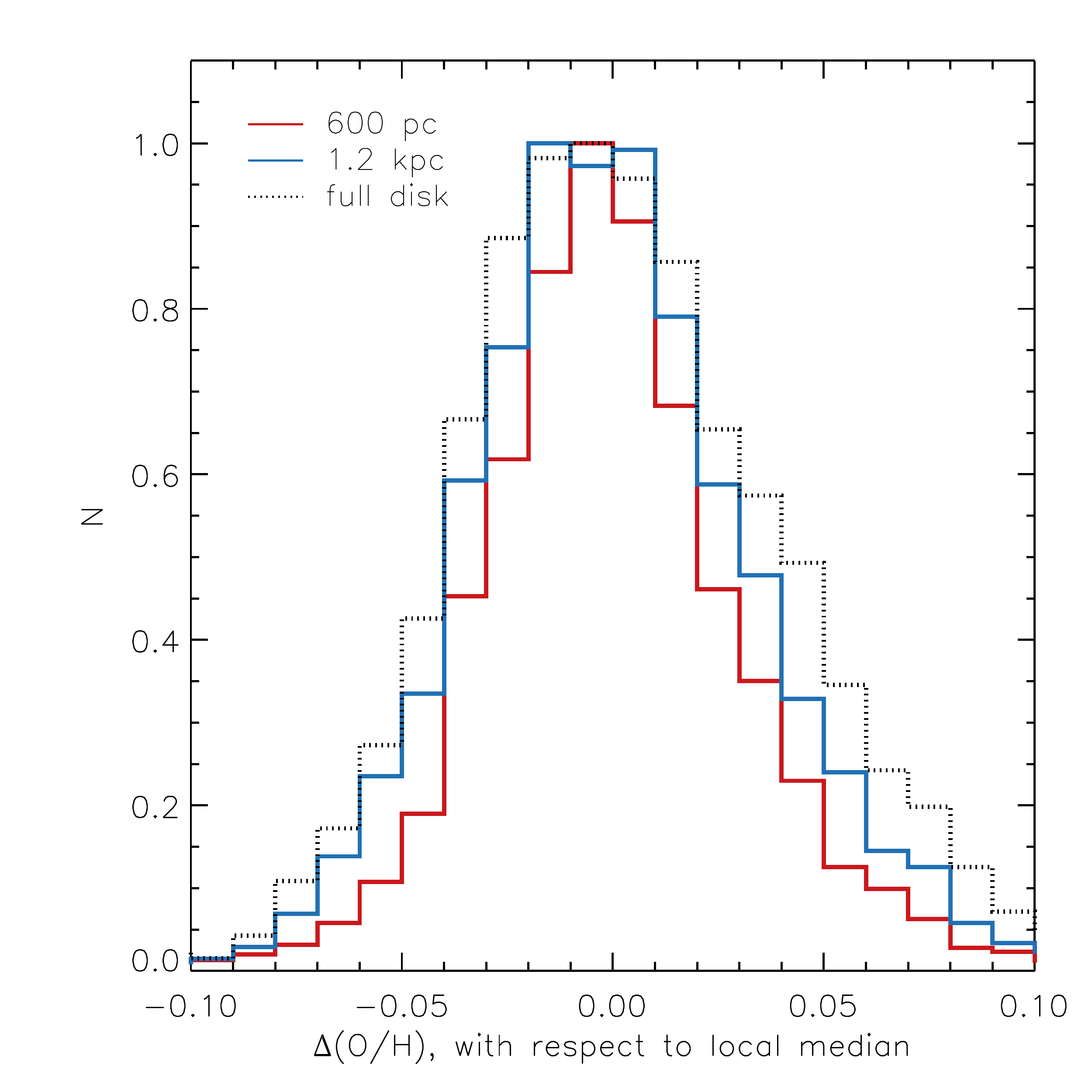}
\caption{Distribution of metallicity offsets from a radial gradient, \doh, at different physical scales. To compare the widths of these distributions, we measure \doh\  with respect to the local median within independent apertures with diameters of 600~pc (red), 1.2~kpc (blue), and globally (dotted) for all galaxies in our sample taken together. The typical (statistical) uncertainty in the measured metallicity is $\sim$0.02 dex.  Metallicities are more homogeneous on smaller scales, and well represented by a Gaussian distribution (Table \ref{tab:results}).
\label{fig:scat_hist}}
\end{figure}

\begin{figure*}
\centering
\includegraphics[width=3in]{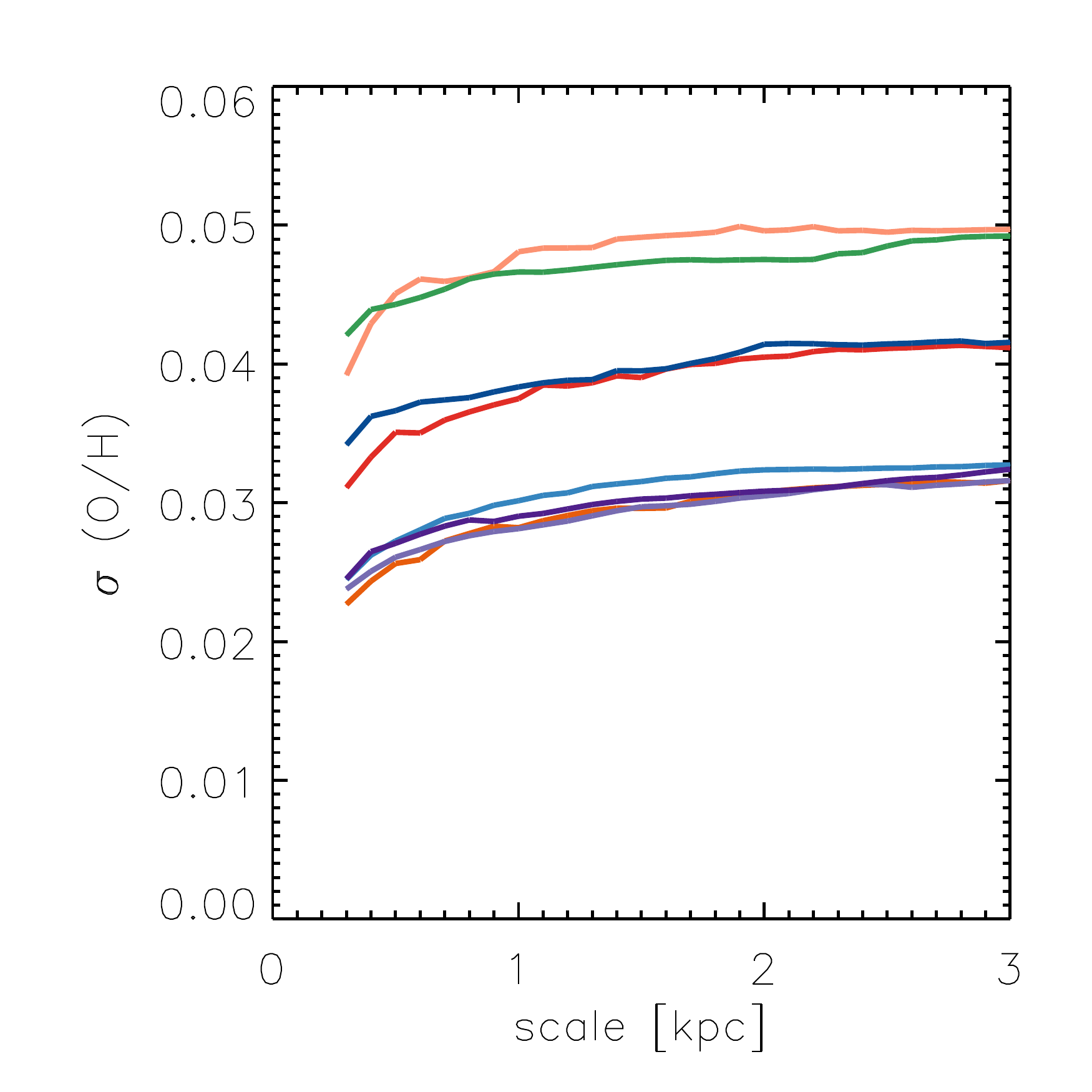}
\includegraphics[width=3in]{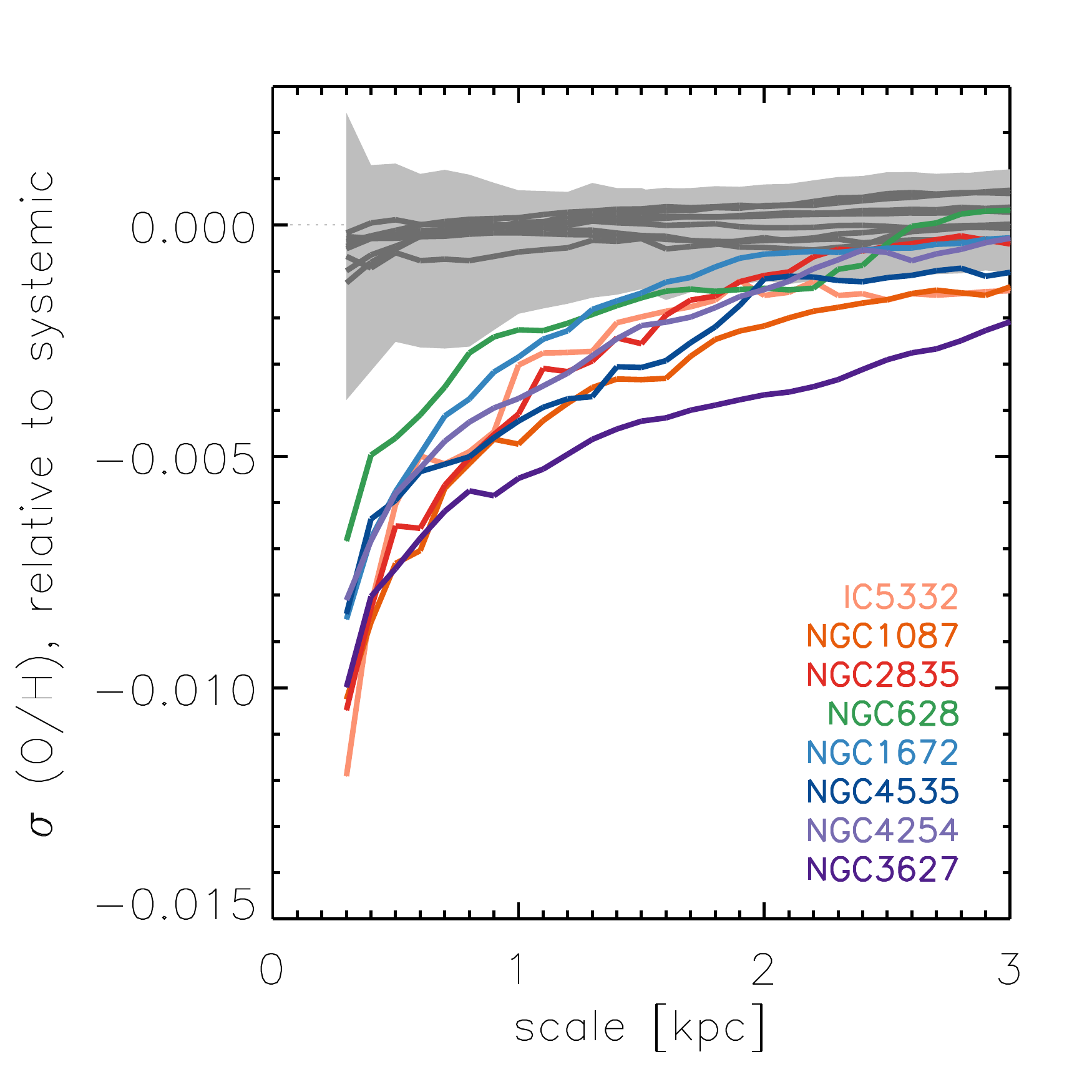}
\caption{Median scatter in \doh\ over a given spatial scale, for apertures centred on \hii\ regions and containing at least 4 \hii\ regions. Here we apply a correction when measuring  $\sigma$  based on Cochran's theorem (assuming a normal distribution of values) to ensure that we achieve an unbiased estimation of the standard deviation. The observed trends (left) and trends in each galaxy measured with respect to the global scatter (right) show consistent homogenization on sub-kpc scales. Galaxy names are sorted by total stellar mass, from low (orange) to high (purple). 
In grey, we test the null hypothesis by randomly shuffling \doh\ 100 times for each galaxy. In dark grey we show the median of each galaxy separately for all realizations, and the light grey band traces our one sigma confidence interval across the sample. In our test of the null hypothesis, we are unable to reproduce the pronounced increase in homogeneity at smallest spatial scales, and find that our signal is detected at the three sigma level.   
\label{fig:scatters3}}
\end{figure*}

\begin{figure*}
\centering
\includegraphics[width=7in]{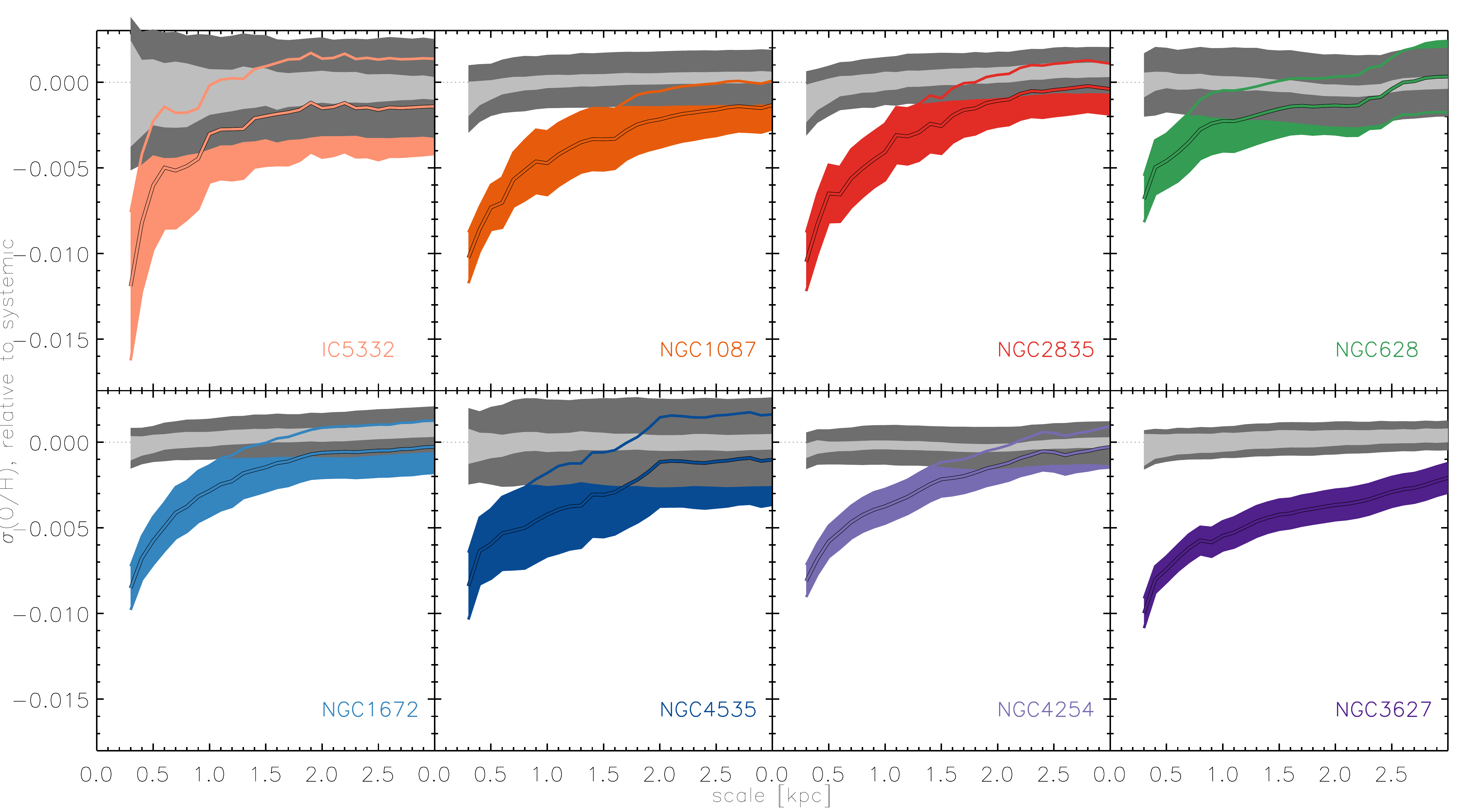}
\caption{Median scatter in \doh\ relative to the global scatter over a range of spatial scales, as in Figure \ref{fig:scatters3}, for each galaxy individually.   The colored range indicates the uncertainty after boot-strapping the error in metallicity over 100 realizations. In grey, we test the null hypothesis by randomly shuffling \doh. The bands show the 1$\sigma$ distribution given 100 different realizations of the shuffling (light grey) and 100 different instances of Gaussian sampling a single randomized realization (dark grey).  IC~5332 has the smallest sample size, and we are unable to disprove the null hypothesis at 3$\sigma$ confidence.  The remaining seven galaxies show increased homogeneity on small $<$1~kpc scales at the 1--3$\sigma$ level.  
\label{fig:scatters3_individual}}
\end{figure*}

From the image atlas presented in Paper I, we see that the metallicity distribution in all galaxies is dominated by a radial gradient.  In this paper we do not include images for all galaxies, but present NGC~4254 as an example in Figure \ref{fig:visualize_new}. The top left panel maps the observed metallicities.  In the top right panel we have subtracted the linear radial metallicity gradient from \hii\ region measurements, and map the variations in metallicity (\doh).  These are generally small (0.03--0.05 dex) when measured globally across the full galaxy disc.  In Paper I, we estimate that the maximum level of enrichment we observe could be caused locally by a 5--10 Myr long period of continuous star formation.  

A striking detail apparent in the maps is that after subtracting a linear radial gradient, neighboring \hii\ regions show very little scatter in \doh. To visualize the size scale over which we observe this increased homogeneity, in Figure \ref{fig:visualize_new} (bottom panels) we calculate the standard deviation in \doh\ (\sigmaoh) at each \hii\ region position over fixed 600~pc, 1.2~kpc and 3~kpc deprojected diameter apertures.  We require each aperture to have at least four \hii\ regions in order to calculate the standard deviation, which holds for 90--95\% of 600~pc deprojected diameter apertures.  When calculating the standard deviation on 600~pc scales, we see connected, $\sim$kpc sized regions with very low scatter (\sigmaoh$<$0.02), consistent with the metallicity measurement uncertainties.  Similar large regions with low scatter (\sigmaoh$<$0.02) are still seen at 1.2~kpc scales, but at 3~kpc scales most of the map is consistent with the global value (\sigmaoh$=0.03{-}0.04$).

\subsection{Global statistics}
In Figure \ref{fig:scat_hist}, we compare the global distribution in \doh\ at different physical scales, across all galaxies in our sample.  To highlight the width of the distribution, for this figure we show \doh\  measured relative to the local median. To minimize repeated sampling of clustered \hii\ regions, we  use independently sampled 600~pc, 1.2~kpc and global apertures. Thus, on local scales, a narrower width in the distribution corresponds to a smaller intrinsic scatter in metallicities. We see directly the reduced scatter in the histograms for the smallest 600~pc scale size, with a minimum observed scatter of $\sim$0.02 dex that is consistent with our quoted errors. 
Statistics relating to these distributions for each galaxy individually are given in Table \ref{tab:results}, including the global scatter (standard deviation), skewness and kurtosis in \doh\ and the average scatter over 600~pc diameter apertures for each galaxy separately.  

\begin{table*}
\footnotesize
\caption{Measured values quantifying the homogeneity and  correlation length scale of the ISM}
\label{tab:results}
\begin{tabular}{lcccccc}
\hline
{Name} &
{Global scatter } &
{Global } &
{Global } &
{Average } &
{50\% } &
{30\% }
\\
{} &
{(standard deviation} &
{skewness} &
{kurtosis} &
{600~pc scatter} &
{Correlation} &
{Correlation}
\\
{} &
{in $\Delta$(O/H))} &
{in $\Delta$(O/H)} &
{in $\Delta$(O/H)} &
{in $\Delta$(O/H)} &
{Scale} &
{Scale}
\\
{} &
{} &
{} &
{} &
{} &
{(pc)}  &
{(pc)} \\
\hline
NGC628 & 0.049 &   0.355 &  -0.365 & 0.040 &  230$\pm$   8 &  360$\pm$  14 \\
NGC1087 & 0.033 &  -0.184 &   0.009 & 0.020 &  380$\pm$   7 &  710$\pm$  54 \\
NGC1672 & 0.033 &   0.193 &   0.058 & 0.021 &  370$\pm$   6 &  760$\pm$  62 \\
NGC2835 & 0.042 &   0.012 &   0.259 & 0.033 &  290$\pm$   2 &  510$\pm$  30 \\
NGC3627 & 0.035 &   0.393 &   0.025 & 0.024 &  370$\pm$   5 &  820$\pm$  79 \\
NGC4254 & 0.032 &   0.454 &   0.028 & 0.022 &  340$\pm$   8 &  740$\pm$  39 \\
NGC4535 & 0.043 &   0.372 &  -0.295 & 0.032 &  290$\pm$   2 &  550$\pm$  35 \\
IC5332 & 0.051 &   0.025 &   0.036 & 0.039 &  270$\pm$   8 &  510$\pm$  77 \\
\hline
\end{tabular}
\end{table*}

We note that in Figure \ref{fig:scat_hist} the global distribution in \doh\ is not symmetric. This is because, as shown in Paper I, the linear radial gradients are fit accounting for the uncertainties in each measurement. The higher metallicity regions are often brighter and hence have correspondingly smaller uncertainties, driving the asymmetry. In this figure, our distribution is shown relative to the median value, which sits at negative \doh, and results in the apparent skewness towards positive values.

\subsection{Trends with physical scale}

To investigate any trends with physical scale, we take circular apertures with increasing diameter around each \hii\ region in order to calculate \sigmaoh\ and plot the median value as a function of scale (Figures \ref{fig:scatters3} and \ref{fig:scatters3_individual}). Here, we have required a minimum of four HII regions be contained in the aperture.  We have chosen this limit as small sample sizes result in a systematic underestimation of the standard deviation. To ensure that we achieve an unbiased estimation of the standard deviation, we apply a correction  based on Cochran's theorem (assuming a normal distributions of values) when calculating \sigmaoh. This corresponds to a correction factor of $\sim$8\% for a sample of four and $<$3\% for a sample of 10. At 300~pc (600~pc) scales we find that $\sim$90\% ($\sim$20\%) of apertures contain 10 or fewer \hii\ regions.   
Figure \ref{fig:scatters3} shows the resulting standard deviation in absolute values (left) and relative to the global scatter (right). 

All galaxies show very similar behavior, with decreased scatter (increased homogeneity) on the smallest sub-kpc scales.  We tested if this could be produced by instrumental effects, and found that the signal persists both within and across different MUSE pointings. Global values of \sigmaoh\ also vary, and the apparent quantization is striking, but we find no obvious correlation with galactic properties such as stellar mass, star formation rate, metallicity, or metallicity gradient (though our sample size is limited). Repeating this analysis with independent, regularly sampled apertures gives quantitatively similar results but with increased scatter (due to more limited statistics). 

As the \hii\ regions are not uniformly distributed across the disc, but often clustered in and around spiral arms, we also test the null hypothesis (that metallicities are uncorrelated) by randomly shuffling all metallicities across the disc and repeating the analysis (in grey).  We repeat this shuffling procedure 100 times for each galaxy, in order to determine the confidence interval of the signal we detect (in light grey). In all galaxies the increased homogeneity on sub-kpc scales is significantly different from the null hypothesis (a randomized distribution). We also perform a similar test, modeling all \hii\ to have a simple linear radial gradient, introducing a scatter equivalent to the global \sigmaoh. This test produces qualitatively similar trends, and is unable to reproduce the observed homogeneity on small spatial scales. 

Figure \ref{fig:scatters3_individual} demonstrates that for each individual galaxy (except IC~5332) the small-scale homogeneity is significantly different from the null hypothesis (that metallicities are uncorrelated), even when taking into account the observational uncertainties. For each galaxy, the colored range indicates the uncertainty after bootstrapping the error in metallicity over 100 realizations.  The light and dark grey bands also confirm the confidence of our detection, including both different realizations of shuffling the metallicities (light grey) and resampling the errors (dark grey).  IC~5332, the least massive galaxy in the sample, shows the least significant trend, but it also has the most limited statistics, with only 223 \hii\ regions detected. The remaining galaxies show significantly increased homogeneity out to scales of 600~pc and larger.

\section{Measuring the correlation scale}
\label{sec:corr}
\begin{figure*}
\centering
\includegraphics[width=7in]{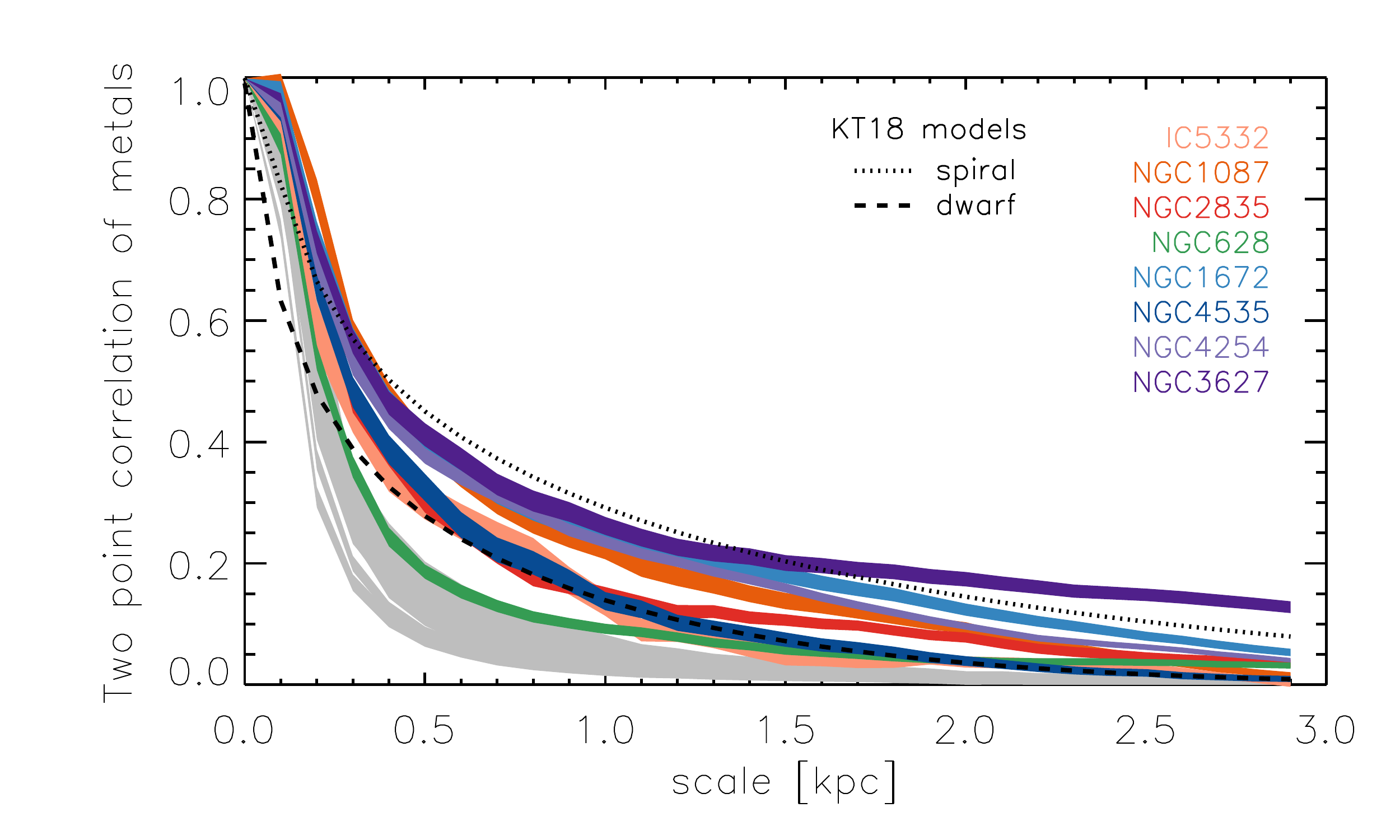}
\caption{Two point correlation of metallicity, \doh\ (after removing the radial gradient) as a function of separation between HII regions, comparing all galaxies. The bands show the 1 sigma distribution given 100 different instances of Gaussian sampling the metallicity uncertainties. The null hypothesis (in grey), that metallicities are uncorrelated, is tested by performing this analysis when all values have been randomized, where the bands show the 1$\sigma$ distribution given 100 different realizations of the shuffling. Most galaxies show correlations at the 30\% level out to nearly kpc scales (see also Figure \ref{fig:corr_individual}). Black lines are from the KT18 \citep{Krumholz2018} model.  The dotted and dashed lines represent the fiducial spiral and dwarf galaxies, respectively, to give a sense of how increased star formation and stellar mass impact the model shape. Galaxy names are ordered by stellar mass, from low (orange) to high (purple). 
\label{fig:corr}}
\end{figure*}

\begin{figure*}
\centering
\includegraphics[width=7in]{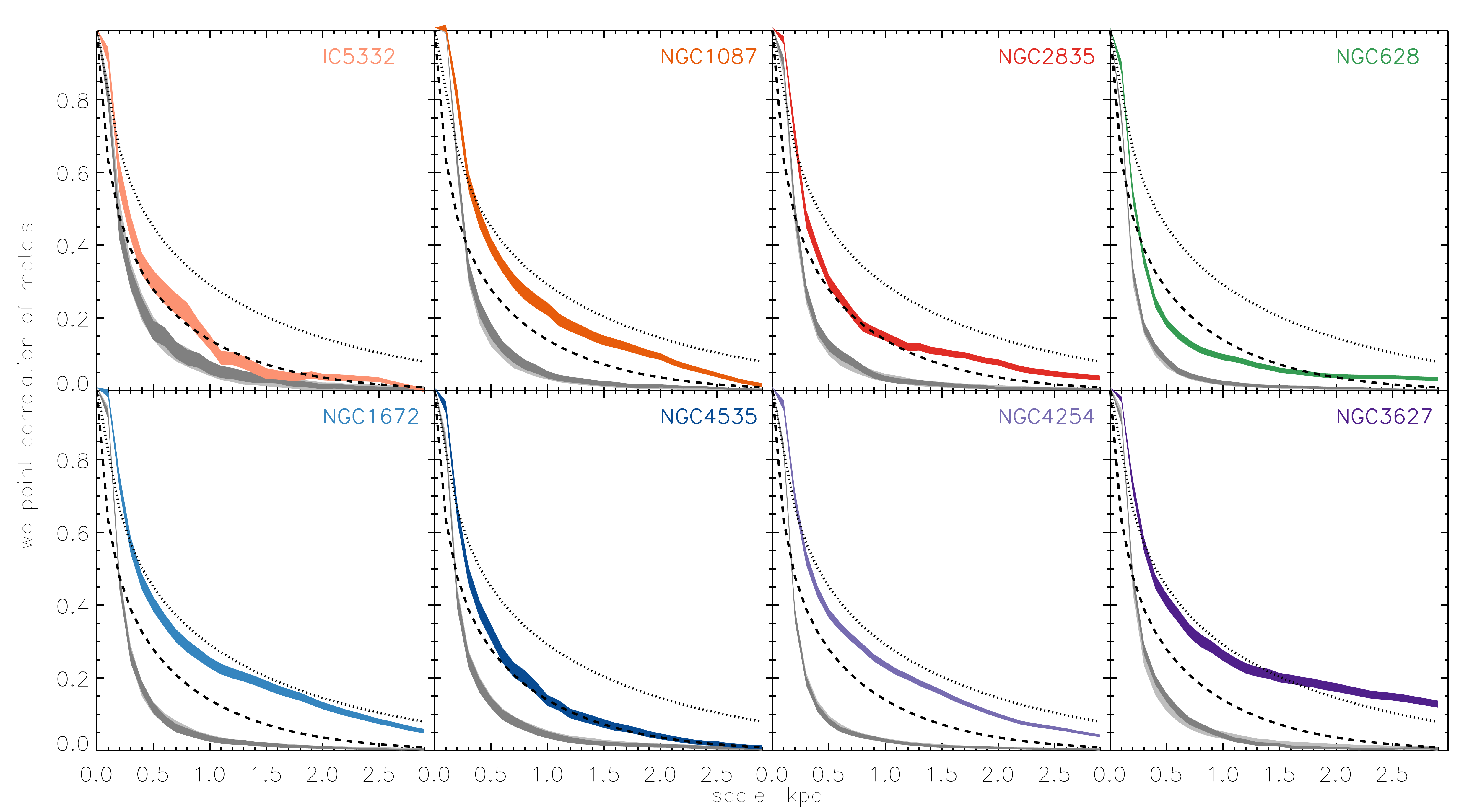}
\caption{Two point correlation of metallicity, \doh\ (after removing the radial gradient) as a function of separation between HII regions, as in Figure \ref{fig:corr}, for each galaxy individually. The colored region indicates the uncertainty estimated as a function of scale  using Bootstrap sampling of the reported uncertainties in the metallicity measurements. The null hypothesis (in grey), that metallicities are uncorrelated, is tested by performing a similar analysis where all values have been randomized. The bands show the 1 sigma distribution given 100 different realizations of the shuffling (light grey) and 100 different instances of Gaussian sampling a single randomized realization (dark grey), which introduce similar levels of uncertainty.  Black lines are from the \citet{Krumholz2018} model. The dotted and dashed represent the fiducial spiral and dwarf galaxies (respectively).  In all galaxies, significant ($>$1$\sigma$) differences are seen across a range of scales, with all showing significantly higher correlations out to  $>$2~kpc scales. Even for NGC~628, which exhibits shorter correlation lengths, this trend is significantly different from the null hypothesis. Galaxies are ordered by stellar mass (low to high; top left to bottom right), with a suggestion of larger correlation scales in the more massive galaxies, roughly consistent with model predictions. 
\label{fig:corr_individual}}
\end{figure*}

\begin{figure*}
\centering
\includegraphics[width=6in]{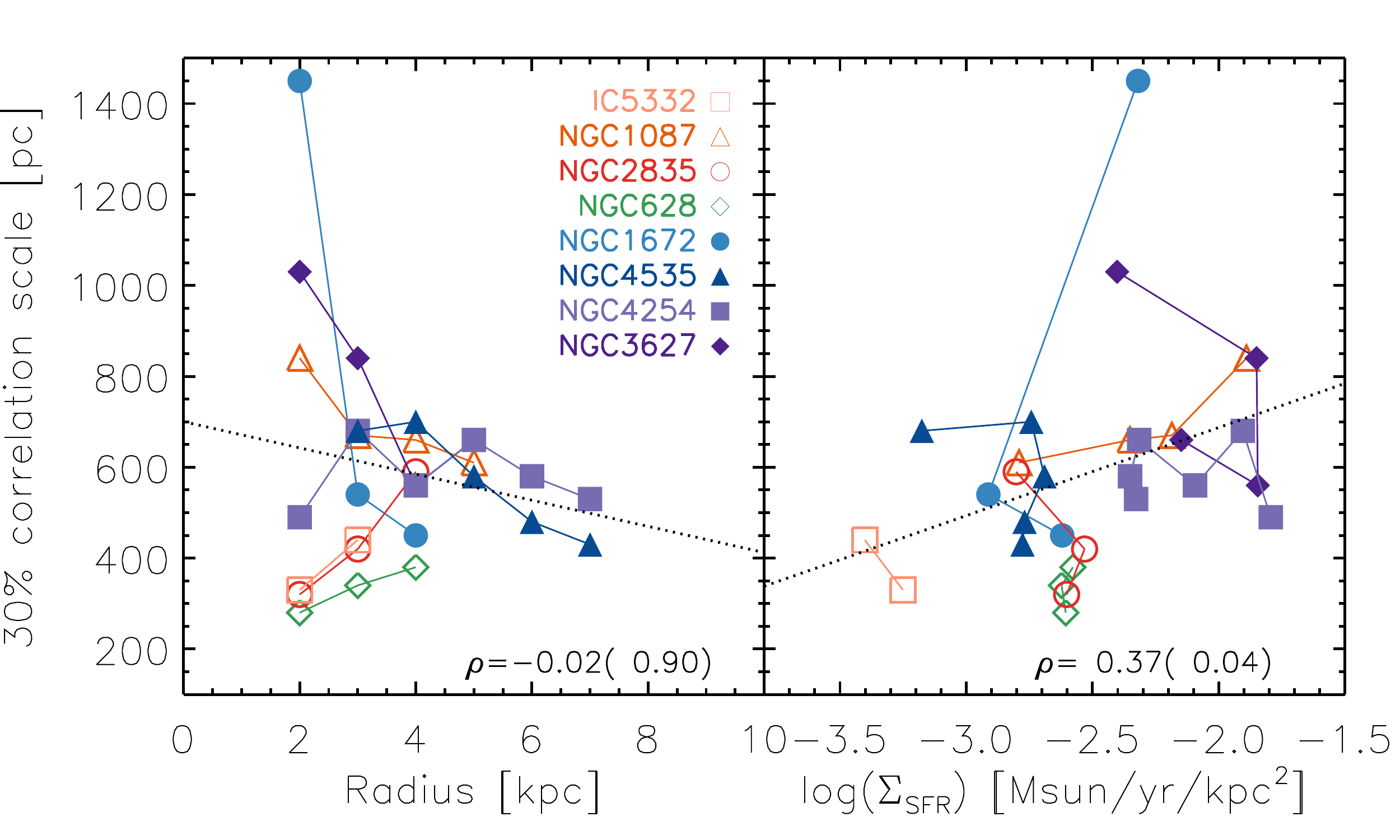}
\includegraphics[width=6in]{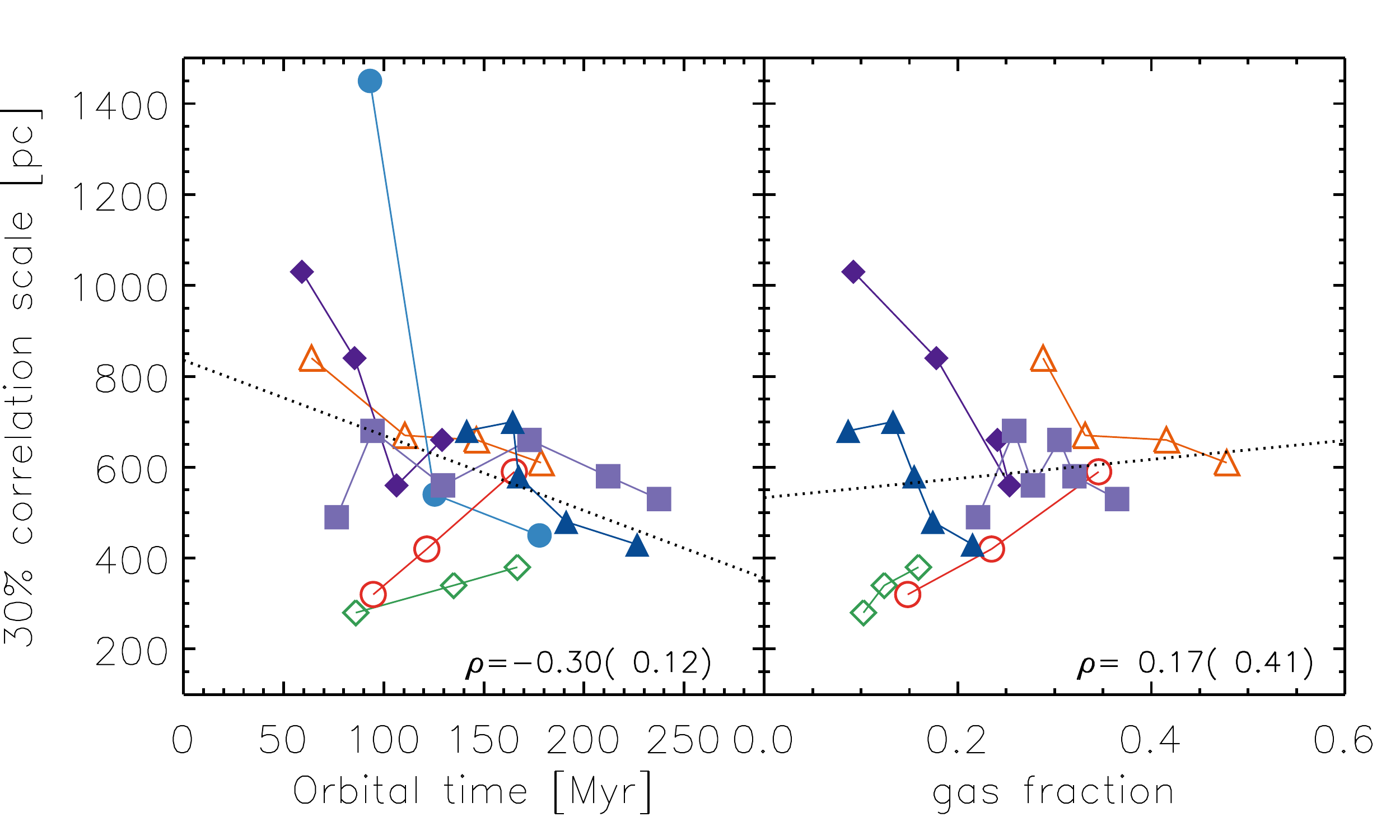}
\caption{The size scale associated with a 30\% correlation (see Figure \ref{fig:corr}) within 2~kpc wide annuli, as a function of radius (top left), $\Sigma_{\rm SFR}$ (top right), orbital time (bottom left), and total gas fraction (bottom right).   
Larger correlation scales are seen for annuli with higher $\Sigma_{\rm SFR}$, and shorter orbital time. However, the Spearman's rank correlation coefficient is not significant (p$\sim$0.1).  No clear trend is seen with gas fraction. Coefficients and significance are shown for each panel in the lower right corners. A linear fit to all points is shown to guide the eye.
\label{fig:pars}}
\end{figure*}

\begin{figure*}
\centering
\includegraphics[width=7in]{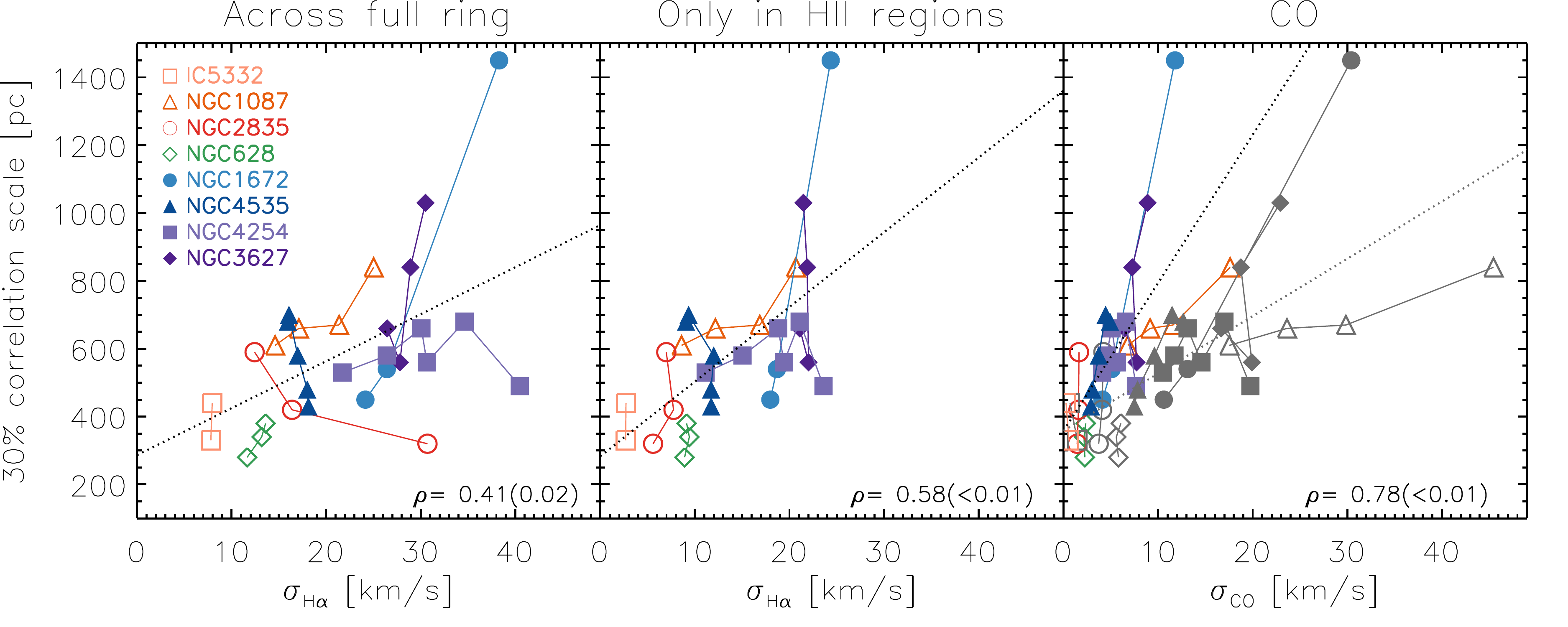}
\caption{The size scale associated with a 30\% correlation (see Figure \ref{fig:corr}) within 2~kpc wide annuli, as a function of the azimuthally averaged ionized gas velocity dispersion.  All measurements are based on pixel-scale (50--100~pc) measurements, and are corrected for instrumental broadening. $\sigma_{\rm H\alpha}$ is the median value from pixels across the full annulus (left) and from only pixels within \hii\ regions (centre). $\sigma_{\rm CO}$ (right) is the median from the second moment map.    In grey, we also show a scaling of the CO velocity dispersions to disk ($\sim$1~kpc) scale heights assuming a power law index of 1/2 \citep{Falgarone2009}. These scaled values are used as input for the \citet{Krumholz2018} model, and are roughly consistent with the H$\alpha$ values. Larger correlation scales are seen for annuli with higher gas velocity dispersion across both ionized gas and molecular gas tracers.  The Spearman's rank correlation coefficient is  significant (p$<$0.01) for all gas phases (lower right corners of each panel). A linear fit to all points is shown to guide the eye.
\label{fig:pars2}}
\end{figure*}

\begin{figure}
    \centering
    \includegraphics[width=3in]{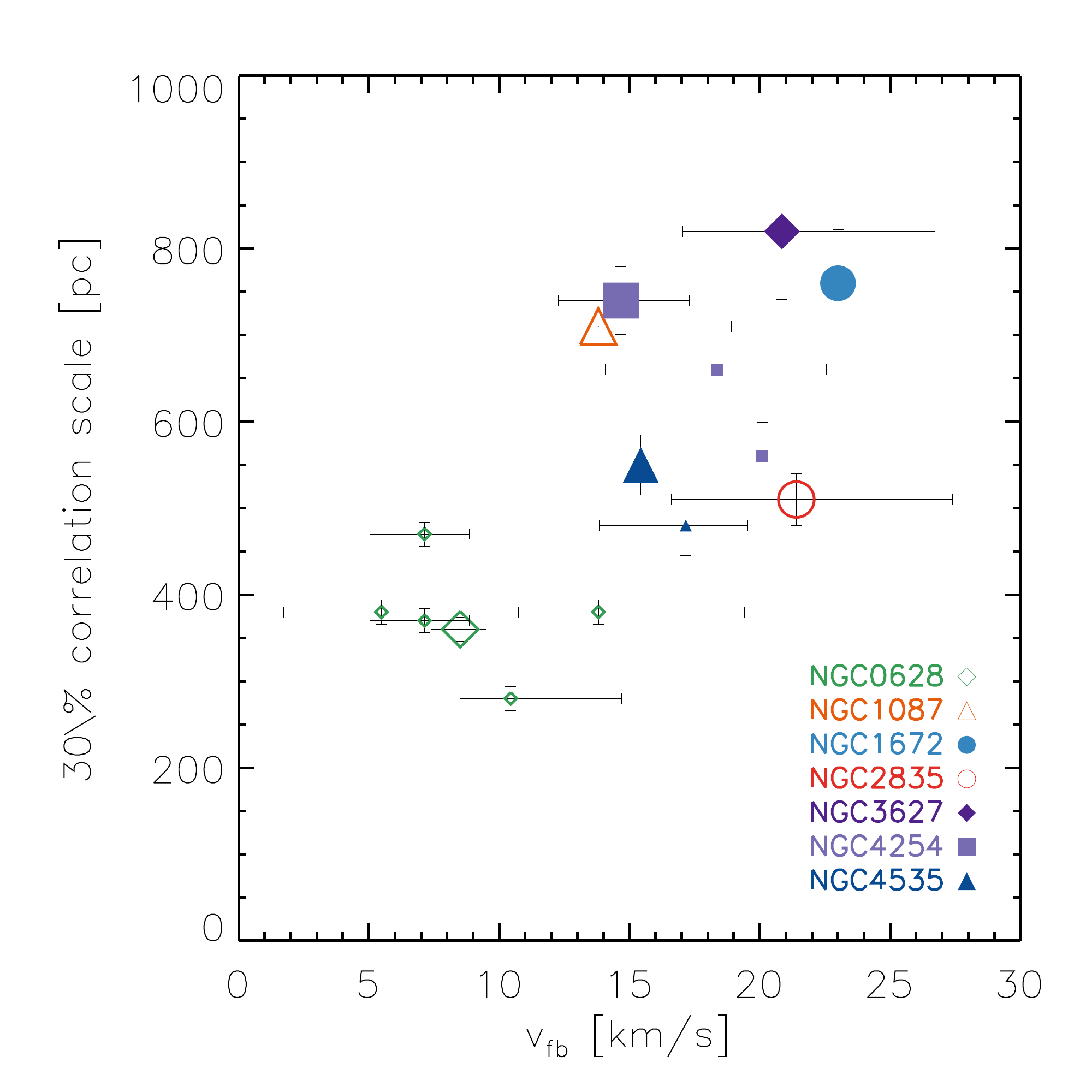}
    \caption{The size scale associated with a 30\% correlation (see Figure \ref{fig:corr}) as a function of the feedback outflow velocity ($v_{\rm fb}$) reported in  \citet{Chevance2020, Kim2020}. The quantity $v_{\rm fb}$ is defined as the ratio of the average CO emission peak radius over the feedback timescale (during which CO and H$\alpha$ are cospatial), as measured using the ``uncertainty principle for star formation'' formalism introduced by \citet{Kruijssen2014} and \citet{kruijssen18}. Global results for the seven galaxies that overlap between samples are shown with large symbols, individual radial bins are shown with smaller symbols. We see a clear correspondence between these two parameters (Spearman's rank correlation coefficient of $\rho=0.78$ with high significance; $p = 0.005$), supporting the indications from Figures \ref{fig:pars} and \ref{fig:pars2} that these mixing scales are set by star formation feedback.}
    \label{fig:vf}
\end{figure}

We now quantify the spatial scale below which the metallicity exhibits reduced scatter (and thus enhanced homogeneity). Consider the vector position of any given \hii\ region in a galaxy, $\mathbf{r_1}$, which provides a measurement at that location of the two dimensional gas-phase oxygen abundance field, $S_X$.  In practice, we measure $S_X(\mathbf{r_1})$ as the offset from the radial gradient (\doh), removing any radial trends from this global analysis. We can then calculate the two point correlation ($\xi$) of the observed gas-phase oxygen abundance as a function of spatial scale (r) for each galaxy as 
 \begin{equation}
 \xi(r) = \left\langle \frac{
 \overline{S_X(\mathbf{r_1}) S_X(\mathbf{r_2})} -
 \overline{S_X}^2}{\overline{\left(S_X - \overline{S_X}\right)^2}} \right\rangle, 
\label{eq:pearson_corr}
 \end{equation}
where $|\mathbf{r_1}-\mathbf{r_2}| \le r$.  The horizontal lines indicate averaging over all \hii\ regions in the galaxy and angle brackets indicate averaging over all choices of $\mathbf{r_1}$. A 100\% correlation is expected at scales of r=0, as each \hii\ region correlates perfectly with itself. The shape of the correlation function traces the scale over which the metal field retains some level of homogeneity. In the following analysis, we parameterize these curves by the scale at which the metallicities remain correlated at the 30\% or 50\% level.  

Figure \ref{fig:corr} compares the correlation functions of all galaxies, while Figure \ref{fig:corr_individual} shows each of our eight galaxies individually.  In all galaxies except NGC~628, the correlation does not drop below 50\% until scales become larger than $\sim$300--400~pc and remains above 30\% out to $\sim$kpc scales. This is a significant result in all galaxies (including NGC~628) when compared to the null hypothesis, where we have randomized $S_X$.  This significance persists when including random sampling of the metallicity uncertainties.  NGC~628 shows shorter correlation scales, but is still significanctly different from the null hypothesis. The global 30\% and 50\% correlation scales for each galaxy are given in Table \ref{tab:results}, and all results presented here are not sensitive to our choice of threshold. 

We compare these observed correlations with the model proposed by \cite{Krumholz2018} for a stochastically forced diffusion model. This model uses an analytic and semi-analytic approach to mimick the diffusion of metals by solving an evolution equation including random (star-formation-driven) injections, combined with linear diffusion as a proxy for turbulent transport.   We find good agreement with their fiducial spiral and dwarf galaxy models (their Figure 5), although their model does not incorporate any structural disc features (e.g.\ spiral arms, bars, or differential rotation).   A more detailed comparison of our results with their model is given in Section \ref{sec:models}.

Breaking each galaxy down into 2~kpc wide (de-projected) radial bins, we explore how the correlation scale changes as a function of local properties that might be expected to drive turbulence and mixing in the ISM.  This annular width is chosen to ensure that most annuli contain at least 100 \hii\ regions and is well matched to the spatial scales available from ancillary data (see Section \ref{sec:ancillary}).  Annuli with less than 80\% of their area contained in the MUSE maps are excluded, as is the inner $r<500$~pc.  We parameterize each annulus by the scale at which the correlation falls below 30\%.  An analysis performed using the 50\% correlation scale gives similar results.

Figure \ref{fig:pars} (top) compares the correlation scale with radius and $\Sigma_{\rm SFR}$.  By the Spearman's rank correlation coefficients calculated over the full sample, radius shows no correlation and $\Sigma_{\rm SFR}$ shows a positive correlation ($\rho\sim0.3$) but with only weak statistical significance ($p\sim0.05$). This figure (bottom) also shows a comparison of correlation scale with orbital time and total gas fraction.  A weak correlation is seen for longer correlation scales at shorter orbital times. However, the Spearman's rank correlation coefficient is not significant ($p\sim0.1$), and no significant correlation is seen with gas fraction. 

Figure \ref{fig:pars2} explores the impact of gas velocity dispersion,  measured in three ways. $\sigma_{\rm H\alpha}$ is the median value from pixels across the full annulus (left) and from only pixels within \hii\ regions (centre). Measurements are corrected for instrumental broadening, but not for thermal broadening. We measure $\sigma_{\rm CO}$ (right) as the median from the second moment CO map, which has already been masked to include only lines of sight with significant CO emission.   This CO emission is concentrated in a thin ($\sim$150~pc) layer within the galaxy mid-plane, a difference reflected by the systematically lower velocity dispersion compared to the $\sim$kpc scales traced by the ionized ISM in H$\alpha$.  In grey, we show how we can scale the CO velocity dispersions to disc ($\sim$1~kpc; \citealt{Levy2019}) scale heights assuming a power law relation between velocity dispersion and spatial scale. Adopting a power law index of 1/2 \citep{Falgarone2009, Klessen2016}, these scaled CO velocity dispersions compare reasonably well with what is measured in H$\alpha$, and retain the strong correlation with the 30\% correlation scale. Across all gas phases, significantly larger correlation scales are seen for annuli with higher gas velocity dispersion with high significance ($p<0.01$).  

Seven of our galaxies have been analysed with the ``uncertainty principle for star formation'' formalism, introduced by \citet{Kruijssen2014} and \citet{kruijssen18}. This technique compares the distribution of spatially resolved ($\sim$100~pc) CO and H$\alpha$ emission peaks, constraining and quantifying the underlying evolutionary timeline associated with the molecular cloud lifetime, the feedback timescale (during which clouds are dispersed), and the (unobscured) \hii\ region lifetime. One parameter measured by this technique is the feedback outflow velocity, defined as the ratio of the average CO emission peak radius over the feedback timescale. This velocity is derived independently of the CO or H$\alpha$ kinematics, and is solely determined using the CO and H$\alpha$ morphology. These measurements are available globally \citep{Chevance2020, Kim2020} as well as in radial bins \citep{Chevance2020b}, and we compare both data sets to our 30\% correlation scales in Figure \ref{fig:vf}. Although the sample size is small, we do observe that galaxies with homogeneity over a larger size scale have larger feedback velocities, with a Spearman's rank correlation coefficient of $\rho=0.78$ and high ($p=0.005$) significance.

\section{Discussion}
\label{sec:discussion}

\begin{figure*}
\centering
\includegraphics[width=7in]{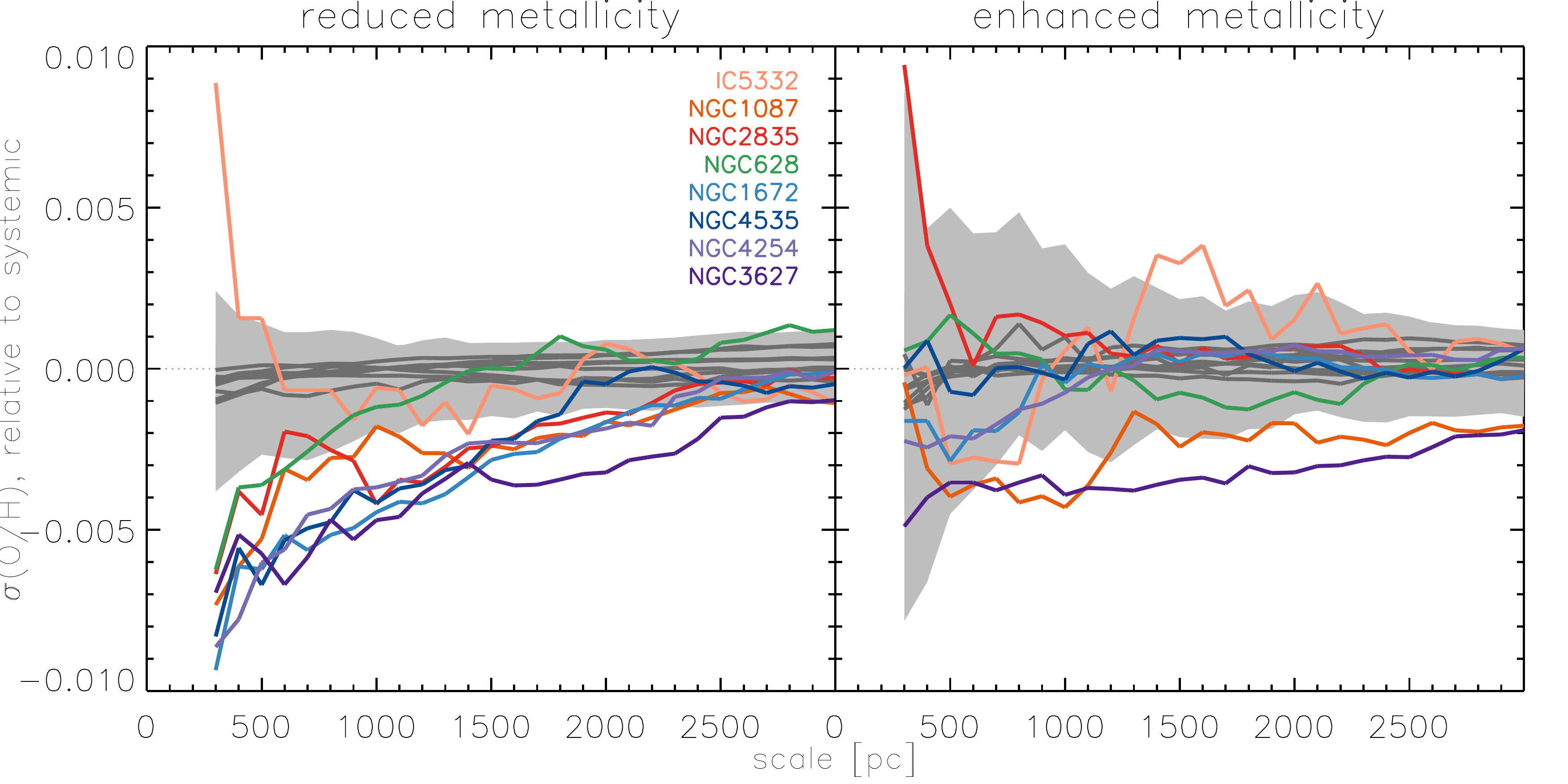}
\caption{ Centring on \hii\ regions with either reduced (left) or enhanced (right) metallicity compared to the local average, we plot the median scatter in \doh\ as a function of aperture scale.  Grey bands show 1$\sigma$ confidence intervals when randomly shuffling the metallicity offsets (over 100 realizations) to test how this analysis appears assuming the null hypothesis (that metallicities are uncorrelated). Dark grey lines are the median of this sample. Regions with reduced abundances show a similar trend for increased homogeneity on small scales as in Figure \ref{fig:scatters3}. Only IC~5332 shows no such trend, but this galaxy also contains the smallest \hii\ region sample size. Regions with enhanced abundances do not show a significant  difference from the null hypothesis.  
\label{fig:hilo2}}
\end{figure*}

\begin{figure}
\centering
\includegraphics[width=3in]{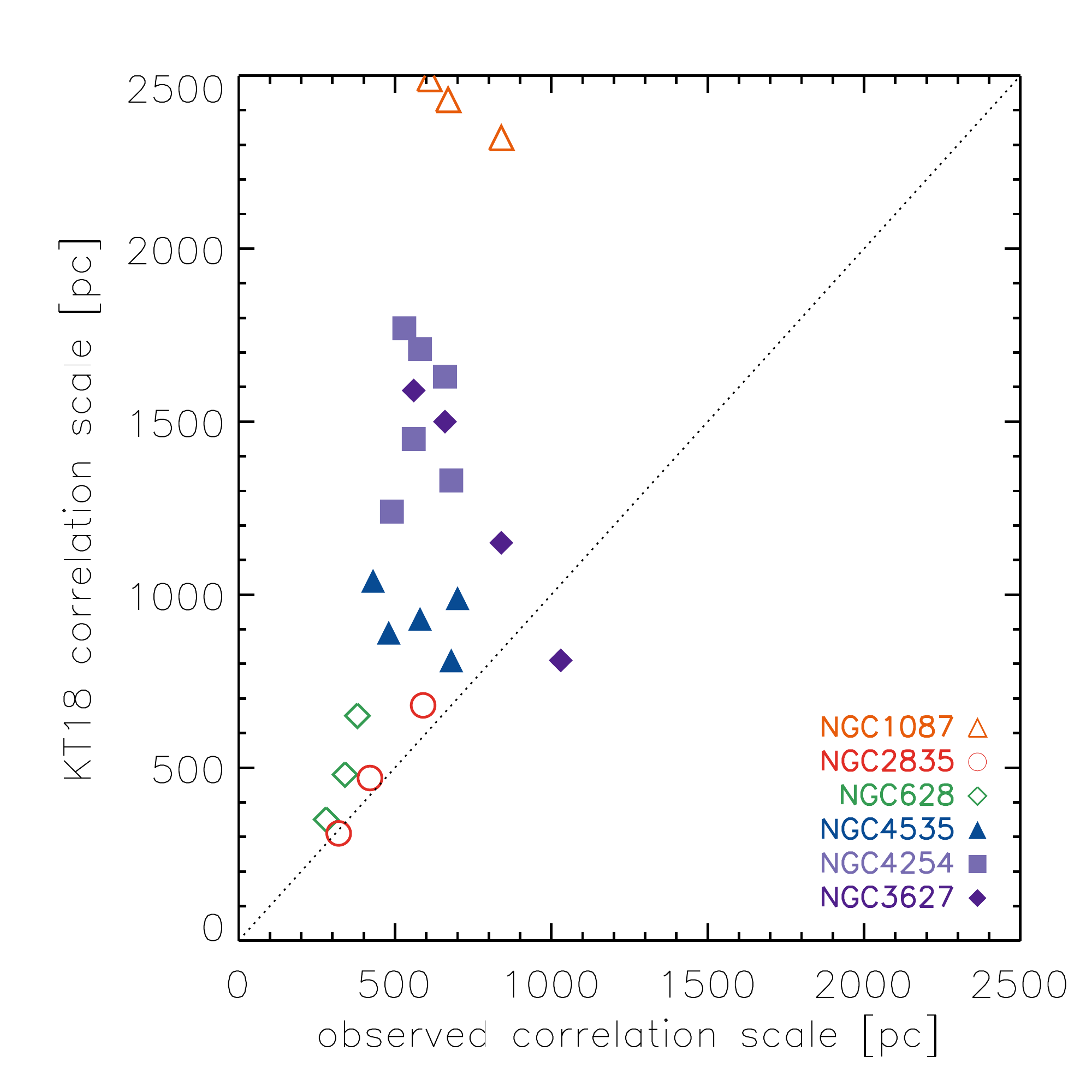}
\includegraphics[width=3in]{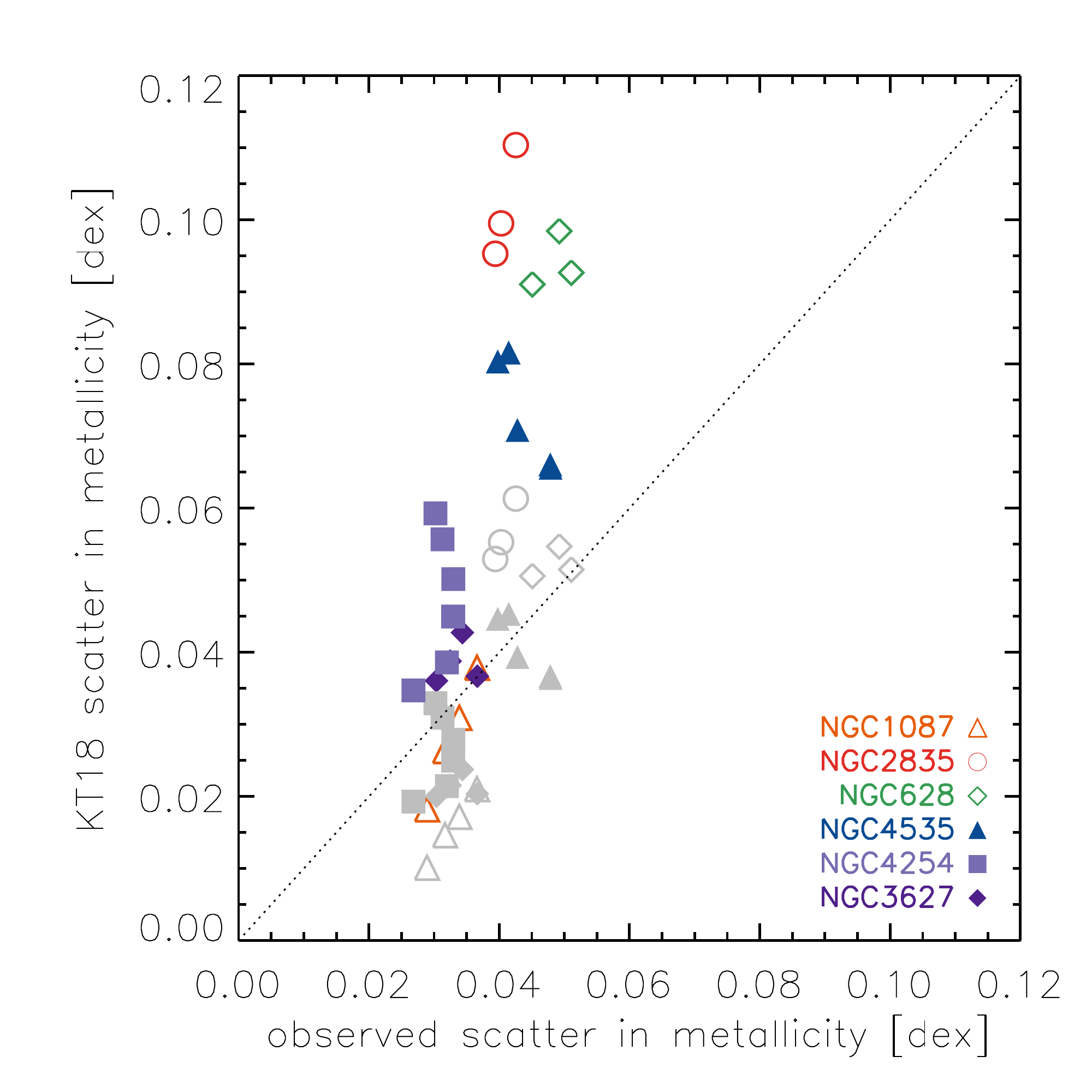}
\caption{Comparison of the predictions from the \citet{Krumholz2018} stochastically forced diffusion model  with our observed values within 2~kpc wide annuli. Measured input parameters for their model in each annulus are shown in Figure~\ref{fig:pars2}. A 1-to-1 line is drawn for comparison.  Galaxies without HI (NGC~1672 and IC~5332) are omitted.  Top: The correlation scale associated with the two point correlation function at the 30\% level.  Only NGC~4254, an interacting galaxy on the Virgo cluster outskirts, falls systematically in disagreement with their model. Bottom: The scatter in metallicity (\sigmaoh). Typical disagreement is by a factor of about two when assuming 50\% of metals are retained in the disc for elements whose primary origin site is Type II SNe.  Increasing this fraction to 90\% (in grey) improves the agreement of the model with observations. 
\label{fig:model}}
\end{figure}

As we report in Paper I, we observe very small systematic scatter (0.03-0.05 dex) globally across all galaxies in our sample after subtracting the radial gradients.  This agrees well with the $\sim$0.05 dex systematic scatter observed by \cite{Sanchez-Menguiano2018} using strong line methods to measure metallicities within 102 spiral galaxies. This level of scatter is much smaller than the large-scale scatter typically measured using direct temperature methods ($\sim$0.1 dex; \citealt{Rosolowsky2008}; \citealt{Croxall2015}; \citealt{Berg2015}; \citealt{Croxall2016}), and it  has been previously shown that strong line methods systematically produce smaller scatter in their radial gradients than direct temperature methods \citep{Arellano-Cordova2016}.  In this work, we report even lower scatter ($\sim$0.02 dex) on the smallest $<$600~pc scales (Table \ref{tab:results}).  This discrepancy between strong line and direct temperature methods is perplexing, and in fact our observed minimum \sigmaoh\ is consistent with our quoted errors.   

While the metallicities presented here are measured assuming the \cite{Pilyugin2016} calibration, we note that all results are qualitatively reproduced with the theoretical \cite{Dopita2016} N2S2 calibration as well as the empirical  \cite{Marino2013} N2 and O3N2 calibrations (see Appendix \ref{appendix}). In addition, the detailed azimuthal trends we reported in Paper I based on strong-line abundances have been confirmed with auroral line electron temperature determinations in one of our galaxies (NGC~1672; \citealt{Ho+2019}), increasing our confidence in the statistical trends we report in this paper.

We measure 30\% correlation scales of 300--800~pc (Table \ref{tab:results}). This indicates the typical physical scales over which mixing in the ISM is most effective.  This has not been well quantified in previous literature. \cite{Sanchez2015} define a `mixing length' based on the dispersion of \hii\ region galactocentric distances with respect to a linear metallicity gradient in NGC~6754.  They measure a 4.6~kpc `mixing length', which parameterizes how far a certain \hii\ region has moved from its expected location based on pure inside-out chemical enrichment. This is much larger than the 30\% correlation scales we report here.  In fact, by our analysis most galaxies appear to have reached their global scatter by $\sim$3~kpc scales.

We note that a measurement of length scales in this context is inextricably linked to the timescales involved in the ongoing pollution of the ISM by stars and supernovae.  In Figure \ref{fig:vf} we observe a strong correlation between our 30\% correlation scales and the feedback outflow velocities reported in \cite{Chevance2020}, which derives from the timescales associated with the dissolution of molecular clouds. This, combined with the correlations with $\Sigma_{\rm SFR}$ and gas velocity dispersion (Figures \ref{fig:pars} and \ref{fig:pars2}) suggests that stellar feedback drives much of the mixing while (presumably) simultaneously contributing to the ongoing enrichment of the ISM.

In this section, we further discuss how the homogeneity measured in Section \ref{sec:homogeneity} and the correlation length measured in Section \ref{sec:corr} can be interpreted within the context of the injection and dilution of metals, as well as in relation to the disc morphology and spiral structure, and finally how this compares with existing diffusion and mixing models.  

\subsection{Injection versus dilution of metals}
The mixing process is responsible for taking material that is enriched within stars and injected through winds and supernovae into the ISM, and redistributing it throughout the disc. The overall efficiency of this process is reflected in the well established mass-metallicity relation \citep{Tremonti2004, Sanchez2019}, that ties the buildup of stellar mass to the buildup of metals in the disc and may lead to the negative radial metallicity gradients measured in almost all nearby galaxies \cite[e.g.][]{magrini2016}. We further explore the trends reported in Section \ref{sec:homogeneity}, focusing on only those \hii\ regions with enhanced abundances relative to the radial gradient ($1\sigma <$ \doh), and only those  \hii\ regions with reduced abundances (\doh $< -1\sigma$). As these are defined to be the 1$\sigma$ outliers  (Figure \ref{fig:scat_hist}), these two samples are roughly equal in size. The use of a fixed threshold for the definition of both samples (e.g. 0.03 dex) produces similar results.

As Paper I showed, within each galaxy the \hii\ regions with enhanced abundances are distributed with wider separation than the \hii\ regions with reduced abundances. In Figure \ref{fig:hilo2}, we reproduce the analysis in Figure \ref{fig:scatters3} but using only these two populations.  Measuring the median \sigmaoh\ as a function of scale for only \hii\ regions with reduced abundances, we recover the same systematic trend for increased homogeneity on small scales. Here, all trends are shown as the offset from the global scatter (Table \ref{tab:results}), and compared with the null hypothesis (that all metallicities are uncorrelated), which we test by randomizing all \hii\ region metallicities. In contrast, the median \sigmaoh\ of \hii\ regions with enhanced abundances show no such increased homogeneity on small scales, particularly in relation to the null hypothesis.    

The stark difference  observed in Figure \ref{fig:hilo2} between these populations indicates that the mechanisms driving homogeneity in the ISM act similarly on \hii\ regions with average abundances and \hii\ regions with decreased abundances relative to the radial gradient.  In contrast, the enriched pockets appear to represent a local phenomenon, and show that the injection of metals is not responsible for the observed chemical homogeneity. This is in good agreement with our hypothesis in Paper I that these enhancements reflect local enrichment produced by 5-10 Myr length episodes of star formation. In this case, the new metals have not yet diffused more broadly into the ISM.

The lengths scales over which we observe homogeneity are similar to the ionized and atomic gas scale heights \citep{ferriere2001}, and hint towards large-scale processes being at work. For example, the persistence of star formation over cosmic time \citep{Prochaska2009} as well as the presence of deuterium in many regions of the Milky Way \citep{linsky2003, lubowich2000} suggests that large-scale accretion of fresh material on to a galaxy must be a common and continuous process \cite[see also][]{sancisi2008}. If the new gas is accreted from the cosmic web \citep{dekel2009} it has considerably lower metallicity than the galactic ISM, and because it falls predominantly on to the outer regions of the disc, the accretion process will steepen the radial metallicity gradient. On the other hand, if the infalling gas is part of a galactic fountain flow \citep{marasco2012,fraternali2017}, then it has been enriched and lifted out of the disc by stellar feedback in the inner regions of the galaxy and typically comes back at larger radii. In this case it will flatten any preexisting metallicity gradient.  Regardless of its origin, the kinetic energy associated with the infalling motion can contribute significantly to the observed turbulence in the galactic ISM \citep{Klessen2010} and  enhance to the global mixing efficiency.  In combination with the spiral driven ``carousel'' model of enrichment and mixing (see Section \ref{sec:spirals} below), this could provide an additional mechanism that facilitates the dilution of gas in the disc while simultaneously driving mixing.

\subsection{Relating disc dynamics and environments with mixing}
\label{sec:spirals}
A similar analysis separating \hii\ region populations by environment (bar, arm, interarm) shows no clear systematic trends. However, the difficulties faced in robustly identifying spiral features make this result difficult to interpret.  Individual case studies \citep{Sanchez-Menguiano2016, Ho2017, Vogt2017, Ho2018} have revealed that spiral structure plays an important role in establishing azimuthal enrichment patterns in the ISM, but in our sample of eight galaxies we find galaxies can vary dramatically, with only half containing subsections of spiral arms with clearly associated azimuthal metallicity gradients (Paper I). As the expected trends depend on a determination of the pattern speed and corotation radius \citep{Spitoni2019}, a full dynamical model of individual galaxies would be needed to reveal trends \citep[see, e.g.\, the case study of  NGC 1672, where clear chemical enrichment is observed along the spiral arm ridge ][]{Ho+2019}, but this is beyond the scope of this paper. 

Since it is clear that the spiral arms play some role in organizing and mixing the ISM, we can interpret our results within the context of two existing simple models for how the spiral structure could be influential. Given that most galaxies show negative radial abundances gradients, radial gas flows driven along spiral  structures could introduce azimuthal patterns in the gas phase abundances \citep{Grand2016,Sanchez-Menguiano2016}.  The mixing induced by spiral-driven streaming motions would be characterized (inside of co-rotation) by a metal-rich trailing edge and a metal-poor leading edge, as relatively metal-rich material is driven outward (along the trailing edge) and relatively metal-poor material is driven inward (along the leading edge).  In this scenario, we would expect \hii\ regions with both enhanced and reduced abundances relative to the radial gradient to symmetrically show similar spatial correlations.  As we observe distinct differences in these two populations, we believe this model is not uniformly supported across our sample.   

\cite{Ho2017} present an alternative ``carousel'' model, where enrichment can proceed slowly in small pockets as gas passes through the interarm region, with enrichment peaking at the spiral arm.  Passage through the spiral arm then triggers large-scale mixing. This homogenization effectively redistributes any enriched pockets of material, resulting in overall lower metallicities on the trailing side of the arm compared to along the spiral arm ridge (where metals have maximally accumulated before redistribution). Given the nearly kpc scale mixing we observe, this could potentially also introduce relatively more pristine gas from the surroundings or from freshly accreted material at large scale heights,  systematically decreasing the abundances.  This is more consistent with our observations, as both the average \hii\ regions and the  \hii\ regions with reduced abundances show increased homogeneity (Figure \ref{fig:hilo2}), and the correlation scale is larger for annuli with higher gas velocity dispersion (Figure \ref{fig:pars2}, left).

\subsection{Comparison with mixing models}
\label{sec:models}
Chemical mixing within the ISM is expected to be driven by a combination of processes operating on different physical scales. On large scales, turbulent diffusion can be driven by supernovae or by thermal instability, aided by the effects of differential rotation \citep{Roy1995, deAvillez2002, Klessen2003, Yang2012, Petit2015}.  Hydrodynamical studies of these effects have shown them to be capable of reproducing the observed level of homogeneity required by the shallow negative metallicity gradients and flattening observed in outer galaxy discs.  However, these focused studies generally do not account for the ongoing injection of enriched material. While they are useful for developing our understanding of the physical processes involved in mixing the ISM, they do not provide predictions that can be directly compared with observations.

More recently, \cite{Krumholz2018} developed a theoretical model of stochastically-forced diffusion that is capable of predicting the multiscale statistics in a disc galaxy metallicity distribution.  Their formalism attempts to balance the stochasticity of injection events (e.g.\ from type Ia and type II supernovae, AGB stars and neutron star mergers) with the mixing driven by interstellar turbulence that acts to homogenize the ISM within a simple uniform disc model.  Their resulting fiducial dwarf galaxy and spiral galaxy models (dashed and dotted lines, Figure \ref{fig:corr}) exhibit very good qualitative agreement with the trends we observe, including the high degree of correlation (30--50\%) observed on sub-kpc scales. Note that the difference between the modeled and observed two point correlation functions at the smallest separation scales reflects our observational limitations, given our \hii\ region minimum separation length of $\sim$100~pc (roughly twice the physical resolution).

As the \cite{Krumholz2018} formalism takes as an input a combination of gas fraction, gas phase velocity dispersion, orbital time, and age of the system (their Equations 109 and 111-113), we can apply their model directly to the annuli shown in Figures \ref{fig:pars} and \ref{fig:pars2} to predice both the correlation scale and the scatter in metallicity.  Here, we assume a fixed system age of 10 Gyr, consistent with their fiducial spiral and dwarf galaxy models, and a fraction of metals retained in the disc  of $f_d$=0.5 \citep{Tumlinson2011}.  Breaking the galaxies down into annuli enables us to better account for some of the radial structure in observed galaxies that is missed in this simple model. 

As shown in Figure \ref{fig:pars2},  velocity dispersions within both the ionized and molecular phases positively correlate with the correlation scale. As our limited instrumental resolution in the ionized phase significantly increases our uncertainty in that measurement, we prefer to base this on the measured CO velocity dispersions ($\sigma_{CO}$).  However, the molecular gas is confined to a thin ($\sim$150~pc) layer, that is not representative of the turbulence at play on ionized disc ($\sim$1 kpc) scale heights. To convert $\sigma_{CO}$ to this larger disc scale, we assume a power law relation between scale length (L) and velocity dispersion as $\sigma \propto L^{1/2}$ \citep{Falgarone2009, Klessen2016}.  These scaled values agree reasonably well with the ionized gas velocity dispersions. Decreasing the target scale height by a factor of two results in slightly better agreement between observed and modeled correlation scales, but worse agreement between observed and modeled scatter in metallicities. Finally, we convert our one-dimensional velocity dispersion ($\sigma_{1D}$, measured along the line of sight) into a three-dimensional velocity dispersion ($\sigma_{3D}$) required as input for the model as $3 \times \sigma^2_{1D}= \sigma^2_{3D}$. 

Figure \ref{fig:model} (top) directly compares the modeled 30\% correlation scales with our observations.  A comparison at 50\% correlation scales shows similar results. In general we see good agreement given the model's lack of large-scale inhomogeneities (such as spiral arms, bars or differential rotation). NGC~1087 and NGC~4254 in particular appear to be systematically poorly modeled, with the correlation scale overpredicted by a factor of three.  NGC~4254 resides in the outskirts of the  Virgo cluster, and shows strong evidence in the \ion{H}{I} distribution of a recent interaction \citep{Vollmer2005}. Perhaps driven by these external dynamical effects, the ionized gas velocity dispersion across the galaxy is particularly high (Figure \ref{fig:pars2}, left), and drives the model towards longer correlation scales.  On the other hand, NGC~3627 is also interacting \citep{Haynes1979}, showing high velocity dispersion and hosting a strong stellar bar. It is unclear how much of the observed dispersion is actually turbulent (and hence would be expected to contribute to mixing) and how much is due to imperfect modelling of the bulk velocity on our resolution scale -- given the high velocities associated with the bar and the substantial inclination.  NGC~3627 is, however, better matched by the \cite{Krumholz2018} model.   

The predicted metallicity dispersions (Figure \ref{fig:model}, bottom) are typically about twice as large and show more variation between galaxies than what is observed, though it also clearly correlated with our observed \sigmaoh.  One free parameter in the \cite{Krumholz2018} model is $f_d$, the fraction of metals retained in the disc.  Based on Milky Way observations, they assume a fiducial value of 0.5. However, they note that the true value could be closer to unity.  Increasing $f_d$ to 0.90 (grey points, Figure \ref{fig:model}) decreases \sigmaoh\ by up to a factor of two, bringing the model into better agreement with our observations. Such a large value of $f_d$ might be plausible locally, but could be difficult to reconcile with the complicated picture of inflows, outflows and gas recycling that regulate galaxy evolution \citep{Tumlinson2017}.

\section{Conclusions}
\label{sec:conclusion}
We examine the statistical distribution of gas phase abundances within the ISM of eight nearby ($D<17$~Mpc) galaxies.  Each galaxy has hundreds to thousands of \hii\ regions with measured 12+log(O/H) based on strong-line metallicity prescriptions, as observed by the PHANGS-MUSE survey (\citealt{Kreckel2019}; Emsellem et al. in prep). After fitting and removing a linear gradient in Paper I, we find rms scatter in residual metallicity of 0.04--0.05~dex. This dispersion becomes even lower (0.02--0.03 dex) when we consider the scatter on local 600~pc scales, indicating a remarkable level of chemical homogeneity.

We apply two-point correlation function statistics to directly measure the 30\% and 50\% correlation scales of the oxygen abundance in each galaxy, both globally and in 2~kpc wide annuli. These size scales correlate most strongly with the gas velocity dispersion, and weakly with SFR surface density.  Our observed statistical correlation lengths are found to be in good agreement with predictions from the stochastically forced diffusion model of \cite{Krumholz2018}, despite the relative simplicity of their disc model. 

\hii\ regions with reduced abundances relative to the radial gradient show consistent and larger scale levels of homogeneity than the \hii\ regions with enhanced abundances, suggesting the uniformity is driven predominantly by  turbulence mixing material from large scales and potentially introducing relatively more pristine material from the halo.  This is consistent with a spiral arm driven `carousel' model of enrichment \citep{Ho2017}, where  gas passing through the interarm region is slowly enriched and metallicities peak at the spiral arm (as shown in Paper I). However, enriched pockets of gas requiring at least $\sim$10 Myr timescales to allow pollution (assuming a continuous star formation history) are observed to survive within this turbulent mixing environment. Overall, the high level of chemical homogeneity over large spatial scales demonstrates that efficient mixing within the ISM is a common feature in spiral galaxies.

\section*{Acknowledgements}

We thank the anonymous referee for their comments which helped improve the clarity of the work. 
KK and FS gratefully acknowledges funding from the Deutsche Forschungsgemeinschaft (DFG, German Research Foundation) in the form of an Emmy Noether Research Group (grant number KR4598/2-1, PI Kreckel). 
SCOG and RSK acknowledge support from the DFG via SFB 881 ``The Milky Way System'' (Project-ID 138713538; sub-projects B1, B2 and B8) and from the Heidelberg cluster of excellence EXC 2181-390900948 ``STRUCTURES: A unifying approach to emergent phenomena in the physical world, mathematics, and complex data'', funded by the German Excellence Strategy. RSK furthermore thanks for funding from the European Research Council via the ERC Synergy Grant ECOGAL (grant 855130). ER acknowledges the support of the Natural Sciences and Engineering Research Council of Canada (NSERC), funding reference number RGPIN-2017-03987. 
FB acknowledges funding from the European
Research Council (ERC) under the European Union's Horizon 2020 research and
innovation programme (grant agreement No.726384/Empire). 
JMDK and MC gratefully acknowledge funding from the DFG through an Emmy Noether Research Group (grant number KR4801/1-1). JMDK, MC, and JJK gratefully acknowledge funding from the DFG through the DFG Sachbeihilfe (grant number KR4801/2-1). JMDK gratefully acknowledges funding from the European Research Council (ERC) under the European Union's Horizon 2020 research and innovation programme via the ERC Starting Grant MUSTANG (grant agreement number 714907).
EW acknowledges support from the DFG via SFB 881 ``The Milky Way System'' (Project-ID 138713538; sub-project P2).
TGW acknowledges funding from the European Research Council (ERC) under the European Union's Horizon 2020 research and innovation programme (grant agreement No. 694343).

This work was carried out as part of the PHANGS collaboration. Based on observations collected at the European Organisation for Astronomical Research in the Southern Hemisphere under ESO programme IDs 094.C-0623(A),  098.C-0484(A), 1100.B-0651(A) and 1100.B-0651(B). 
This paper makes use of the following ALMA data:
ADS/JAO.ALMA\#2012.1.00650.S, 
ADS/JAO.ALMA\#2015.1.00925.S, 
ADS/JAO.ALMA\#2015.1.00956.S, 
ADS/JAO.ALMA\#2017.1.00392.S, 
%
ALMA is a partnership of ESO (representing its member states), NSF (USA) and NINS (Japan), together with NRC (Canada), MOST and ASIAA (Taiwan), and KASI (Republic of Korea), in cooperation with the Republic of Chile. The Joint ALMA Observatory is operated by ESO, AUI/NRAO and NAOJ. The National Radio Astronomy Observatory is a facility of the National Science Foundation operated under cooperative agreement by Associated Universities, Inc.

\section*{Data Availability}
The data underlying this article are available in \citet{Kreckel2019}.



\bibliographystyle{mnras}
\bibliography{mpia} 








\appendix

\section{Alternate metallicity prescriptions}
\label{appendix}
We reproduce two key figures from this paper (Figures \ref{fig:scatters3} and \ref{fig:corr}) using three additional strong-line metallicity prescriptions, also presented in Paper I (Appendix C). Due to the wavelength coverage of MUSE, a limited number of calibrations are available.

Our preferred calibration (Scal; \citealt{Pilyugin2016}) is based on an empirical comparison with \hii\ regions where direct temperature measurements are available, and relies on a combination of three emission line ratios. This relatively high dimensionality in the prescription helps break the degeneracy between ionization parameter and metallicity variations. For comparison, we show here results using two other empirical calibrations (O3N2 and N2; \citealt{Marino2013}), which include fewer diagnostic ratios and thus are less able to account for the influence of ionization parameter variations.  We further compare with a photo-ionization model based calibration (D16; \citealt{Dopita2016}), which by design is also intended to break the degeneracy between ionization parameter and metallicity. In Paper I we found the Scal and D16 calibrations to show very similar qualitative trends in azimuthal variations, though D16 produced significantly higher global scatter in metallicity variations, \sigmaoh.  

\begin{figure*}
\centering
\includegraphics[width=5.5in]{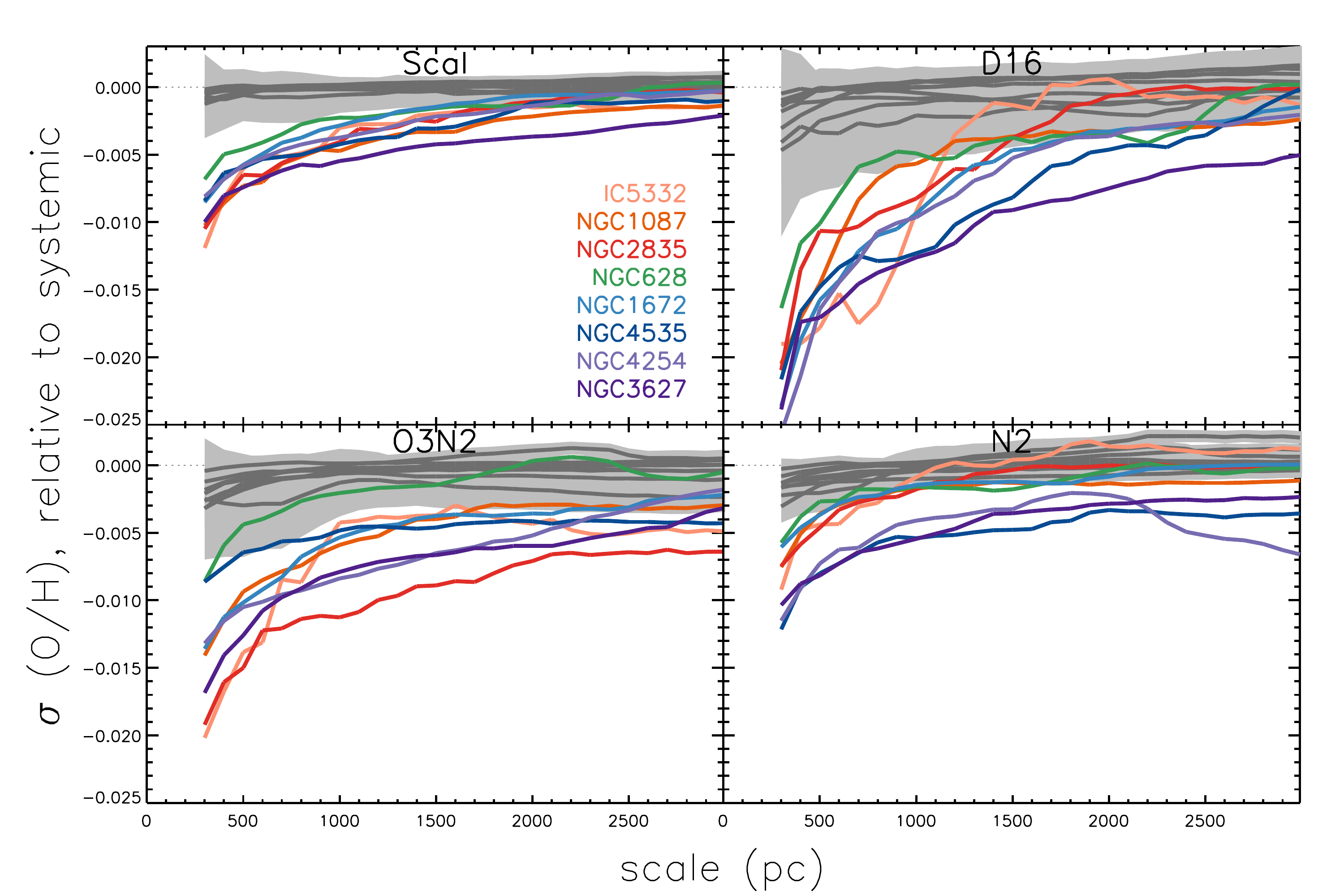}
\caption{Revisiting Figure \ref{fig:scatters3}, using different metallicity calibrations, to show the median scatter in \doh\ over a given spatial scale, for apertures centered on \hii\ regions. Trends in each galaxy are measured with respect to the systemic scatter. Our preferred empirical calibration (Scal; \citealt{Pilyugin2016}) shows similar qualitative trends as the three other calibrations explored in Paper I. This includes photo-ionization model based calibrations (D16; \citealt{Dopita2016}) and alternate empirical prescriptions (O3N2, N2; \citealt{Marino2013}). 
In all calibrations, we see increased homogeneity on small, scales, in excess of what is predicted by the null hypothesis (in grey).
\label{fig:appendix_scatter}}
\end{figure*}

\begin{figure*}
\centering
\includegraphics[width=5.5in]{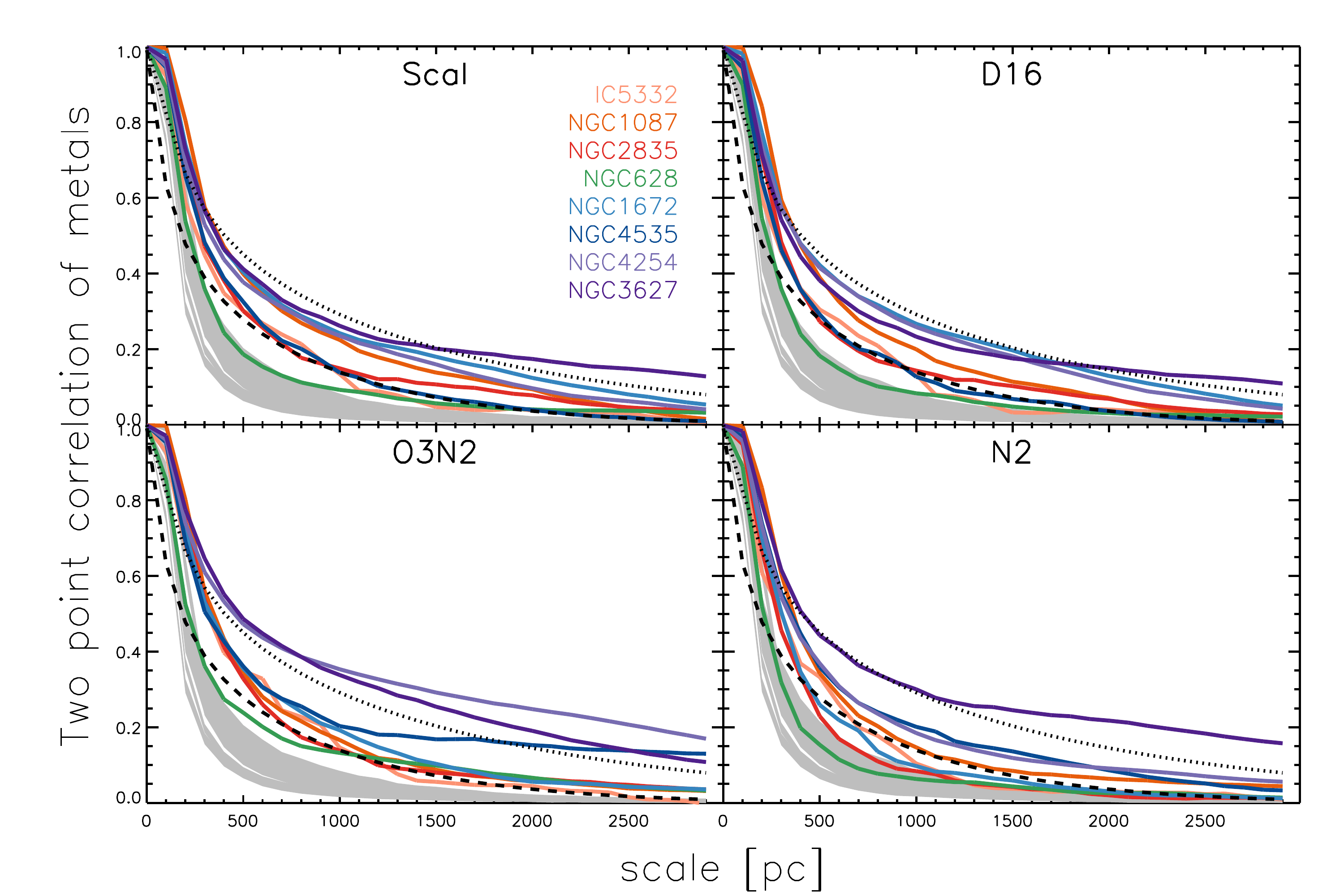}
\caption{Revisiting Figure \ref{fig:corr}, using different metallicity calibrations, to show the two point correlation function of metals as a function of scale. Our preferred empirical calibration (Scal; \citealt{Pilyugin2016}) shows similar qualitative trends as the three other calibrations explored in Paper I. While the photo-ionization model based calibration D16 \citep{Dopita2016} shows larger global scatter (\sigmaoh; Paper I, Table 3), quantitatively the correlation scales agree quite well with the results reported here.  The empirical O3N2 and N2 calibrations \citep{Marino2013} are unable to account for variations in ionization parameter, and both show on average shorter correlation scales (but still in excess of the null hypothesis, in grey).   
\label{fig:appendix_corr}}
\end{figure*}

As is seen in Figures \ref{fig:appendix_scatter} and \ref{fig:appendix_corr}, the Scal-based results presented in this paper show good qualitative agreement with all other calibrations. 

\vspace{4mm}

\noindent {\it
$^{1}$Astronomisches Rechen-Institut, Zentrum f\"{u}r Astronomie der Universit\"{a}t Heidelberg, M\"{o}nchhofstra\ss e 12-14, 69120 Heidelberg, Germany\\
$^{2}$Max Planck Institut f\"{u}r Astronomie, K\"{o}nigstuhl 17, D-69117 Heidelberg, Germany\\
$^{3}$The Observatories of the Carnegie Institution for Science, 813 Santa Barbara Street, Pasadena, CA 91101, USA \\
$^{4}$ Departamento de Astronom\'{i}a, Universidad de Chile, Casilla 36-D, Santiago, Chile\\
$^{5}$ Universit\"{a}t Heidelberg, Zentrum f\"{u}r Astronomie, Institut f\"{u}r theoretische Astrophysik, Albert-Ueberle-Str. 2, 69120 Heidelberg, Germany \\
$^{6}$ Universit{\"a}t Heidelberg, Interdisziplin{\"a}res Zentrum f{\"u}r Wissenschaftliches Rechnen, Im Neuenheimer Feld 205,  69120 Heidelberg, Germany \\
$^{7}$International Centre for Radio Astronomy Research, University of Western Australia, 7 Fairway, Crawley, 6009, WA, Australia\\
$^{8}$4-183 CCIS, University of Alberta, Edmonton, Alberta, Canada\\
$^{9}$Argelander-Institut f\"{u}r Astronomie, Universit\"{a}t Bonn, Auf dem H\"{u}gel 71, 53121 Bonn, Germany\\
$^{10}$Centro de Astronom\`{i}a (CITEVA), Universidad de Antofagasta, Avenida Angamos 601, Antofagasta, Chile\\
$^{11}$Department of Physics \& Astronomy, University of Wyoming, Laramie WY, USA\\
$^{12}$IPAC, California Institute of Technology, Pasadena, CA 91125, USA\\
$^{13}$European Southern Observatory, Karl-Schwarzschild-Stra{\ss}e 2, D-85748 Garching bei M\"{u}nchen, Germany\\
$^{14}$Univ.\ Lyon, Univ.\ Lyon1, ENS de Lyon, CNRS, Centre de Recherche Astrophysique de Lyon UMR5574, F-69230 Saint-Genis-Laval France\\
$^{15}$Research School of Astronomy and Astrophysics, Australian National University, Weston Creek 2611, Australia\\
$^{16}$Department of Astronomy, The Ohio State University, 140 West 18th Ave, Columbus, OH 43210, USA\\
$^{17}$Sydney Institute for Astronomy, School of Physics, A28, The University of Sydney, NSW, 2006, Australia\\
$^{18}$Sterrenkundig Observatorium, Universiteit Gent, Krijgslaan 281 S9, B-9000 Gent, Belgium\\
$^{19}$Departamento de F\'{i}sica Te\'{o}rica, Universidad Aut\'{o}noma de Madrid, Cantoblanco, E-28049 Madrid, Spain\\
$^{20}$Center for Astrophysics and Space Sciences, Department of Physics, University of California, San Diego, 9500 Gilman Drive, La Jolla, CA 92093, USA\\
$^{21}$Max-Planck Institut f\"ur Extraterrestrische Physik, Giessenbachstra\ss e 1, 85748 Garching, Germany
}

\bsp	
\label{lastpage}
\end{document}